\documentclass[5p,times]{elsarticle}

\usepackage{amssymb}
\usepackage{graphicx}
\usepackage{subcaption}
\usepackage{pifont}

\usepackage{array}
\usepackage{xurl}
\usepackage[utf8]{inputenc}
\usepackage{textcomp}
\usepackage{multirow}
\usepackage{caption} 
\usepackage{booktabs} 
\usepackage{algorithm}
\usepackage{algpseudocode}
\usepackage{amsmath}
\usepackage{booktabs}
\usepackage{tablefootnote}
\usepackage{listings}
\usepackage{url}
\usepackage{xcolor}
\usepackage{tabularx}  
\usepackage{wrapfig}  
\usepackage[switch]{lineno}
\usepackage{microtype}

\journal{Elsevier}

\begin{document}

\begin{frontmatter}



\title{ContractShield: Bridging Semantic-Structural Gaps via Hierarchical Cross-Modal Fusion for Multi-Label Vulnerability Detection in Obfuscated Smart Contracts}

\author[inst1,inst2]{Minh-Dai Tran-Duong}\ead{22520183@gm.uit.edu.vn}
\author[inst1,inst2]{Nguyen Hai Phong}\ead{22521088@gm.uit.edu.vn}
\author[inst1,inst2]{Nguyen Chi Thanh}\ead{22521350@gm.uit.edu.vn}
\author[inst1,inst2,inst3]{Doan Minh Trung}\ead{trungdm@uit.edu.vn}
\author[inst4]{\\Tram Truong-Huu}\ead{truonghuu.tram@singaporetech.edu.sg}
\author[inst1,inst2,inst3]{Van-Hau Pham}\ead{haupv@uit.edu.vn}
\author[inst1,inst2,inst3]{Phan The Duy}\ead{duypt@uit.edu.vn}

\affiliation[inst1]{organization={Information Security Laboratory, University of Information Technology},
            city={Ho Chi Minh city},
            country={Vietnam}}

\affiliation[inst2]{organization={Vietnam National University Ho Chi Minh City},
            city={Hochiminh City},
            country={Vietnam}}

\affiliation[inst3]{organization={VNU-HCM Information Security Center},
            city={Hochiminh City},
            country={Vietnam}}
\affiliation[inst4]{organization={Infocomm Technology Cluster, Singapore Institute of Technology},
            country={Singapore}}

\begin{abstract}
Smart contracts are increasingly targeted by adversaries employing obfuscation techniques such as bogus code injection and control flow manipulation to evade vulnerability detection. Existing multimodal methods often process semantic, temporal, and structural features in isolation and fuse them using simple strategies such as concatenation, which neglects cross-modal interactions and weakens robustness, as obfuscation of a single modality can sharply degrade detection accuracy. To address these challenges, we propose ContractShield, a robust multimodal framework with a novel fusion mechanism that effectively correlates multiple complementary features through a three-level fusion. Self-attention first identifies patterns that indicate vulnerability within each feature space. Cross-modal attention then establishes meaningful connections between complementary signals across modalities. Then, adaptive weighting dynamically calibrates feature contributions based on their reliability under obfuscation. For feature extraction, ContractShield integrates (1) CodeBERT with a sliding window mechanism to capture semantic dependencies in source code, (2) Extended long short-term memory (xLSTM) to model temporal dynamics in opcode sequences, and (3) GATv2 to identify structural invariants in control flow graphs (CFGs) that remain stable across obfuscation. Empirical evaluation demonstrates ContractShield's resilience, achieving a $89\%$ Hamming Score with only a 1–3\% drop compared to non-obfuscated data. The framework simultaneously detects five major vulnerability types with $91\%$ F1-score, outperforming state-of-the-art approaches by $6-15\%$ under adversarial conditions.
\end{abstract}



\begin{keyword}
Vulnerability Detection \sep Smart Contract \sep Graph Neural Networks \sep Multimodal Learning \sep Code Obfuscation
\end{keyword}

\end{frontmatter}


\section{Introduction}
To support the evolution of decentralized applications, smart contracts have become indispensable. Their ability to autonomously enforce business logic across trustless environments has enabled innovations in finance, governance, and supply chain. However, the growing complexity and criticality of these contracts have simultaneously elevated the risks associated with latent security vulnerabilities \cite{chu2023survey}. This has led to significant security incidents, such as the DAO hack in 2016~\cite{gemini_dao_hack}, where a recursive call vulnerability allowed an attacker to siphon over \$60 million worth of Ether. The dForce hack in 2023~\cite{wanjiku2023dforce} caused a loss of \$3.64 million across both the Optimism and Arbitrum chains, highlighting the ongoing threat posed by vulnerable smart contracts.

Although automated vulnerability detection approaches have been proposed across various paradigms, including symbolic execution (e.g., Mythril), fuzzing (e.g., ContractFuzzer~\cite{jiang2018contractfuzzer}), formal verification (e.g., ZEUS~\cite{kalra2018zeus}), and graph-based reasoning (e.g., MadMax~\cite{grech2018madmax}) — have shown promise, they suffer from critical limitations, including path explosion, shallow code coverage, and high reliance on hand-crafted rules ~\cite{wu2024comprehensive}. Furthermore, most existing benchmark datasets consist of overly simplistic and synthetic contracts, lacking the structural complexity found in real-world deployments~\cite{zhang2020source}. As a result, these tools often fail to demonstrate reliable performance when faced with complex or obfuscated contracts~\cite{wu2024comprehensive}. For instance, Zhang \textit{et al.}~\cite{zhang2023bian} introduced BiAn, the first obfuscation tool for Ethereum smart contracts. It significantly increases contract complexity through techniques such as syntax rewriting and control flow transformation, thereby exposing the limitations of existing detection approaches. Machine learning and deep learning-based approaches rely on unimodal representations \cite{vidal2024vulnerability}, focusing solely on a single source of information (e.g., source code-based~\cite{qian_auto_squen_mol, zhang2022cbgru}, bytecode-based~\cite{vu2023enhancing, tong2024enhancing}, or graph-based~\cite{cheong2024gnn, luo2024scvhunter}), leading to the failure in detection of diverse vulnerabilities, as they tend to focus on known patterns or signatures~\cite{upadhya2024quadracode}. Recent efforts such as VulnSense by J. Hu et al. \cite{hu2025enabling} have leveraged Generalized Zero-Shot Learning (GZSL) to classify unseen vulnerability types based solely on textual descriptions, mitigating the dependency on labeled samples. However, such approaches primarily operate on high-level semantic abstraction and lack the capacity to model modality-specific structural signals inherent in smart contracts.

Recently, multimodal learning has garnered increasing attention in the field of smart contract vulnerability detection~\cite{lian2024universal, jie2023novel, li2023smart, deng2023smart, duy2025vulnsense}. By integrating multiple representations of a contract (e.g., source code, bytecode, control/data flow graphs), these approaches aim to capture complementary semantic and structural information, thereby enhancing detection robustness. However, most of these approaches rely on simple fusion strategies, such as feature concatenation or shallow convolutional operations, which often fail to model complex interdependencies across modalities, such as Hymo \cite{khodadadi2023hymo}, VulnSense by Duy et al. \cite{duy2025vulnsense}, and the work of Deng et al. \cite{deng2023smart}. These methods treat each modality as an independent information source, ignoring the intricate cross-modal relationships that may be crucial to detect subtle or deeply embedded vulnerabilities. However, different modalities often convey asymmetric information—some may encode high-level logic (e.g., source code), while others reflect low-level execution semantics (e.g., bytecode or control flow). This semantic imbalance makes it difficult for naïve fusion mechanisms to produce coherent and representative joint features.

Obfuscation techniques, such as control flow flattening or instruction substitution, often distort the structural and semantic alignment across modalities, rendering traditional fusion methods ineffective in preserving meaningful patterns for vulnerability detection. Consequently, one of the significant challenges in multimodal vulnerability detection is how to effectively extract and align salient features across heterogeneous modalities, especially when analyzing real-world smart contracts that are often obfuscated or structurally complex. Addressing this challenge requires a more sophisticated fusion strategy that captures both intra-modal specificity and inter-modal complementarity. To address these pressing challenges, we propose a novel hierarchical cross-modal fusion mechanism that facilitates fine-grained interaction between modalities through a combination of self-attention, cross-modal attention, and adaptive weighting. This approach allows the model to dynamically align and integrate information from heterogeneous sources, mitigating the issues of modality imbalance and redundancy and improving the model’s ability to detect subtle vulnerabilities. Building on this fusion strategy, we introduce ContractShield, a robust and scalable multimodal framework for multi-label vulnerability detection in Ethereum smart contracts. In particular, ContractShield is designed with strong resilience to adversarial obfuscation. It integrates three complementary views of smart contract behavior: (1) a semantic view using sliding-window-enhanced CodeBERT for contextual representation from source code (SC), (2) a temporal view using xLSTM \cite{beck2024xlstm} to capture opcode dynamics (OP), and (3) a structural view using GATv2 to model CFGs derived from bytecode.

In summary, we make the following contributions:

\begin{itemize} 
    \item We present ContractShield, a multimodal framework for multi-label vulnerability detection in Ethereum smart contracts. The framework jointly models three complementary views of a contract, including source code, opcode sequence, and CFG. These modalities are integrated through a hierarchical fusion mechanism combining self-attention, cross-modal attention, and adaptive weighting to balance their contributions and improve resilience to obfuscation.
    \item We construct a comprehensive obfuscation benchmark suite for evaluating smart contract vulnerability detectors. The suite incorporates realistic transformations such as control-flow reordering, operator substitution, and bogus-code injection, enabling systematic assessment of each modality and its fused representation under adversarial code modifications.
    \item Extensive experiments on four widely used datasets, including SoliAudit \cite{liao2019soliaudit}, SmartBugs \cite{ferreira2020smartbugs}, CGT Weakness \cite{di2023consolidation}, and DAppScan \cite{zheng2024dappscan}. ContractShield consistently outperforms state-of-the-art baselines on clean data, improving Hamming Score by 7–14\% on SoliAudit-SmartBugs, 12–18\% on CGT Weakness, and 20–25\% on DAppScan. Under obfuscation, the model maintains strong accuracy with only modest performance degradation (1.22\%–3.42\%), demonstrating its robustness against structural and semantic perturbations.
\end{itemize}

The remainder of this paper is organized as follows: Section~\ref{sec:related_work} reviews related literature. Section~\ref{sec:methodology} details our proposed approach. Section~\ref{sec:experiments} describes the experimental setup and analyzes the results. Next, Section \ref{sect_Discussion} presents a comprehensive discussion of the results, highlighting their significance, practical implications, and possible threats to validity prior to the conclusion. Finally, Section~\ref{sec:conclusion} concludes the paper and outlines future directions.
\section{Related Work}
\label{sec:related_work}

Machine learning approaches for smart contract vulnerability detection have evolved significantly, spanning both unimodal and multimodal strategies \cite{wang2025review, crisostomo2025machine, bresil2025deep}. We organize our discussion into two primary categories: unimodal methods that rely on single data representations, and multimodal approaches that integrate multiple feature types for enhanced detection robustness.

\subsection{Unimodal smart contract vulnerability detection}
Existing studies in this category predominantly rely on a single type of feature and operate within a single modality of data for smart contract vulnerability detection and classification. These approaches can be categorized into three types: source code-based, bytecode-based, and graph-based methods. 

For source code-based methods, P. Qian et al. \cite{qian_auto_squen_mol} transformed Solidity source code into sequential representations and assembled contract snippets to detect reentrancy vulnerabilities, though their evaluation focused on only a single vulnerability type. Similarly, Yuan et al. \cite{yuan2023optimizing} proposed a transfer-learning-based approach that trains a Transformer encoder on multi-modality representations, including source code, intermediate representation, and assembly code, supported by a Bug Injection framework to improve smart contract vulnerability detection.

For bytecode-based methods, D. Vu \cite{vu2023enhancing} leveraged the pre-trained SecBERT model to extract latent semantic information from bytecode, followed by a Multi-Layer Perceptron (MLP) for multi-label vulnerability detection. V. Tong et al. \cite{tong2024enhancing} similarly combined bytecode representations with SecBERT and a multi-label classification strategy to identify multiple vulnerability types.

More recent research expands bytecode analysis with structural and transformer-based modeling. COBRA \cite{li2025interaction} integrates semantic context by converting bytecode into Static Single Assignment (SSA) form and inferring function signatures through SRIF when ABI metadata is unavailable, improving robustness under incomplete information. VASCOT \cite{balci2025examining} employs transformer architectures with a sliding-window mechanism to analyze long opcode sequences, capturing structural patterns resilient to obfuscation strategies such as padding or reordering. LLM-based systems such as LLM-SmartAudit \cite{wei2025advanced} and SmartGuard \cite{ding2025smartguard} also operate primarily at the bytecode level, applying high-level reasoning over low-level representations through prompting, retrieval, and chain-of-thought inference. However, these models are sensitive to syntactic perturbations and representation obfuscation \cite{yu2022bytecode}, which can degrade prediction accuracy.

For graph-based methods, ByteEye \cite{yang2026byteeye} constructed CFGs directly from Ethereum bytecode and applied graph neural networks (GNNs) to detect multiple vulnerabilities using structural invariants. Nguyen et al. \cite{nguyen2023mando} introduced MANDO-HGT, which builds heterogeneous contract graphs from both source code and bytecode to represent inter-function calls and control-flow relationships. Cheong et al. \cite{cheong2024gnn} transformed opcode sequences into CFGs and employed Sent2Vec to embed node features. These representations were then fed into a GNN for classification, demonstrating the effectiveness of graph-centered learning for vulnerability identification. J. Cai et al \cite{cai2024fine} formulated fine-grained vulnerability detection as a node-classification problem over code graphs, where each node corresponds to an AST subtree, and proposed a syntax-sensitive GNN to identify vulnerable statements.

\subsection{Multi-modal smart contract vulnerability detection}
To overcome the inherent limitations of unimodal approaches, recent studies have explored multimodal representations to construct richer and more robust feature spaces for detecting smart contract vulnerabilities. D. Han et al. \cite{HAN2025130619} introduced MKDD-Vul, a lightweight multi-modal knowledge distillation framework that integrates CFG representations, grayscale image features, and opcode sequence information to capture diverse vulnerability patterns across complementary modalities. W. Jie et al. \cite{jie2023novel} proposed a multi-view learning framework that simultaneously leverages three modalities: source code, bytecode, and CFGs. Their experiments on multiple supervised detection tasks demonstrate that combining all three modalities significantly outperforms models using only one or two data sources. Similarly, W. Deng et al. \cite{deng2023smart} adopted a multimodal decision-fusion strategy that incorporates source code, opcode sequences, and CFG information, showing that fusing signals from multiple modalities consistently yields more accurate detection than unimodal baselines. Extending this direction, J. Li et al. \cite{li2023smart} extracted cross-modal features spanning the full lifecycle of a smart contract. By combining both static and dynamic characteristics and employing deep learning architectures such as GCN and BiLSTM, their framework achieved strong performance in vulnerability detection. More recently, T. Wang et al. \cite{wang2025tmf} proposed TMF-Net, a multimodal vulnerability detection framework that fuses contract graphs, bytecode text, and code images through a multi-scale Transformer fusion network. By extracting modality-specific features using GCNs, BiLSTMs, and CNNs, and integrating them via a multi-head attention–based multiscale fusion module, TMF-Net learns fine-grained semantic representations of smart contracts, achieving notable improvements over existing multimodal approaches.

Multi-modal approaches offer clear advantages: they provide more comprehensive feature representations, improve robustness, and capture complex interdependencies that are often invisible within any single modality. However, most existing multimodal methods rely on relatively simple fusion mechanisms such as feature concatenation or shallow convolutional operations, which limit their ability to model intricate cross-modal interactions. As a result, the shared and complementary patterns across modalities are only partially utilized. Furthermore, many methods depend on decision-level fusion \cite{deng2023smart}, where predictions from individual modalities are combined via averaging or voting. Although effective in mitigating noise, such strategies overlook deeper cross-modal relationships essential for precise vulnerability identification.

\begin{figure*}[!ht]
\centering
\includegraphics[width=0.8\textwidth]{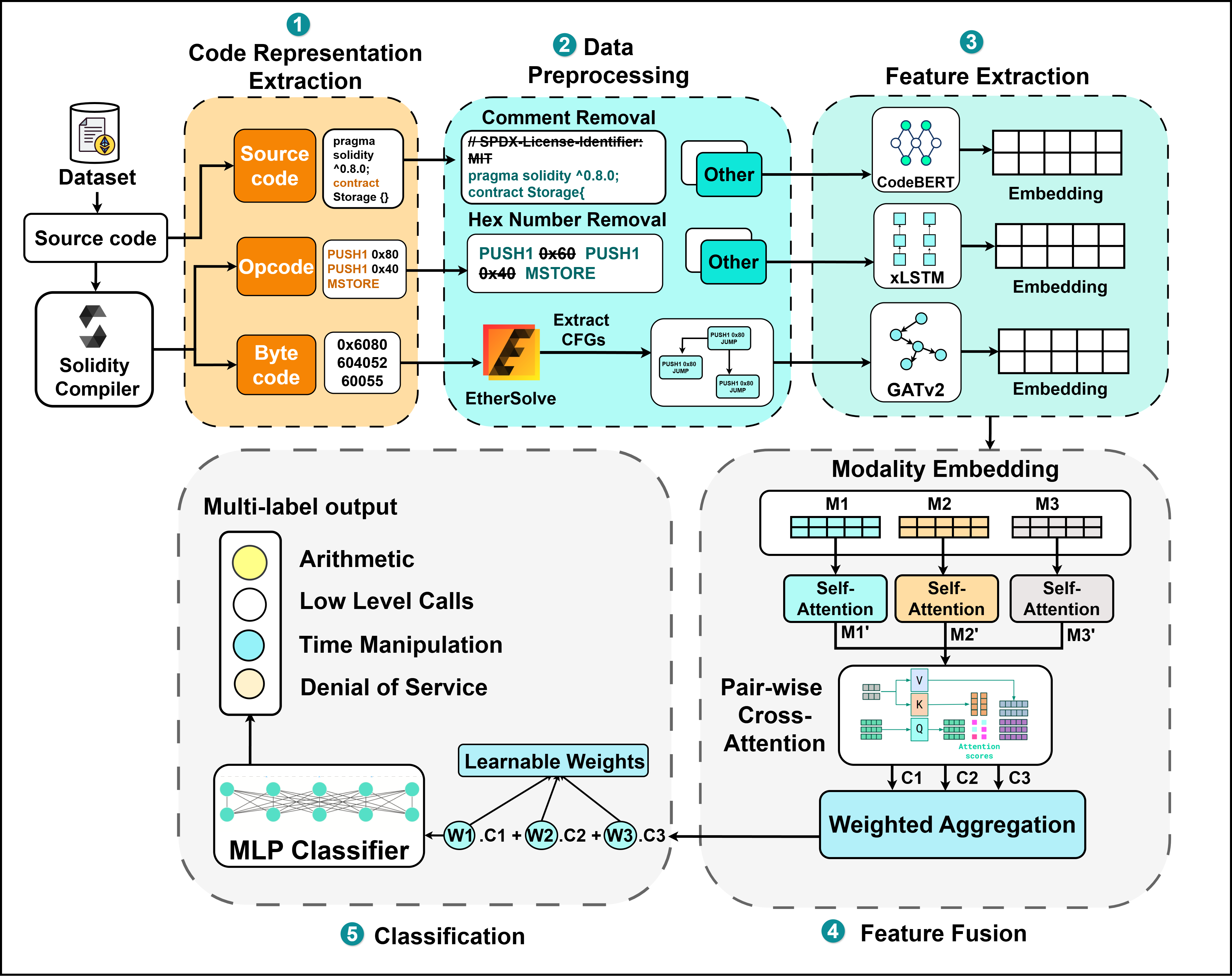}
\caption{The overview structure of the ContractShield framework.}
\label{fig:model_pipeline}
\end{figure*}

\section{Methodology} 
\label{sec:methodology}
This section presents ContractShield, our deep multimodal fusion framework for smart contract vulnerability detection. Our approach implements a comprehensive pipeline that combines multiple code representations to achieve robust vulnerability identification. The framework extracts and processes complementary representations of smart contracts, leveraging specialized neural architectures for each modality before fusing them through a novel hierarchical attention mechanism. This design enables more accurate detection of various vulnerability types while maintaining resilience against obfuscation techniques.

\subsection{Overall framework architecture}
\label{subsec:arch}

ContractShield employs a multi-stage pipeline designed to analyze smart contracts comprehensively from multiple perspectives. As illustrated in \textbf{Figure \ref{fig:model_pipeline}}, our approach consists of five key stages:

\begin{enumerate}
    \item \textbf{Code representation extraction:} We extract three complementary representations from each smart contract: Solidity source code, opcode sequences, and bytecode-derived CFGs. Each representation provides unique insights into contract behavior and potential vulnerabilities.
    
    \item \textbf{Data preprocessing:} Each representation undergoes tailored preprocessing to enhance learning efficiency, including normalization, comment removal, and instruction simplification to preserve semantic integrity while reducing noise.
    
    \item \textbf{Feature extraction:} Specialized neural architectures process each representation: a transformer-based model for source code, a sequential model for opcodes, and a graph neural network for CFGs, capturing unique vulnerability patterns across modalities.
    
    \item \textbf{Hierarchical feature fusion:} Our novel hierarchical fusion mechanism integrates features from all representations through a three-level fusion approach, combining intra-modality and inter-modality information effectively.
    
    \item \textbf{Classification:} The fused representation feeds into a classifier that performs multi-label classification to identify specific vulnerability types.
    

\end{enumerate}

This architecture enables ContractShield to leverage the complementary nature of different contract representations, where vulnerabilities may manifest distinctly across modalities, improving detection accuracy and robustness.

\subsection{Code representation extraction}
\label{subsec:stage1}

\subsubsection{Opcode}
We retrieve the opcode sequence using the Solc compiler, which provides a structured format of opcode instructions essential for analyzing smart contract behavior. These opcodes represent a low-level execution view of the contract, capturing critical operations that may reveal vulnerabilities missed by high-level source code analysis. For example, issues like improper handling of low-level calls (e.g., \texttt{CALL} or \texttt{DELEGATECALL}) often surface at the opcode level but may be obfuscated in Solidity code.

\subsubsection{Control flow graph}
\begin{figure}[!ht]
    \centering
    \includegraphics[width=0.5\textwidth]{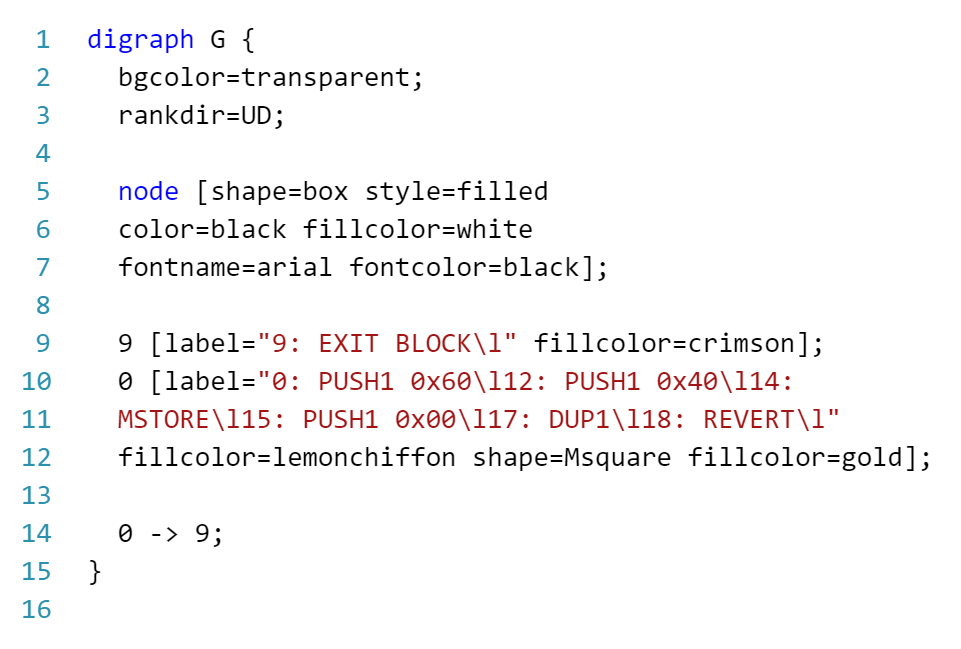}
    \caption{Control flow graph representation of a .dot file.}
    \label{fig:cfg_dot}
    \vspace{-2ex}
\end{figure}

We first extract the runtime bytecode using the Solc compiler and then use EtherSolve \cite{contro2021ethersolve} to generate the CFG from this bytecode in a .dot file format, as shown in \textbf{Figure \ref{fig:cfg_dot}}. The bytecode-level CFG is crucial for detecting complex vulnerabilities, such as reentrancy or access control flaws, which typically emerge through specific control flow patterns. For example, reentrancy vulnerabilities are often manifested in the sequencing and branching of function calls, a behavior that can only be effectively captured by analyzing the bytecode's control flow. While opcode analysis can reveal individual low-level calls, the CFG offers a broader structural perspective, enabling more accurate vulnerability detection by combining both opcode and bytecode analyses.

During the process of representing the program’s CFG, extracting information from nodes and edges is an important step in building a graph representation model for deep neural networks. For the GAT model mentioned earlier, the PyTorch Geometric library requires input in the form of node features and edge indices with labels. In a CFG, nodes typically represent code blocks or contract states, while edges indicate the control links between nodes. 

\textit{Node Extraction}: Using PyGraphviz\footnote{\url{https://pygraphviz.github.io/}}, we parse the .dot file to identify all nodes and their associated opcode sequences. For example, in Figure~\ref{fig:cfg_dot}, node 0 contains the raw opcode sequence PUSH1 PUSH1 MSTORE PUSH1 DUP1 REVERT, representing an initialization or error-handling block, while node 9 is labeled EXIT\_BLOCK, marking a termination point. The opcode sequence of each node is embedded in a fixed-dimensional vector using CodeBERT, which we fine-tune to preserve semantic and syntactic relationships between the EVM instructions.

\textit{Edge Extraction}: The CFG’s edges are extracted as a sparse adjacency matrix in COO (coordinate) format, represented as a tensor of shape (2, N), where N is the number of edges. Each column in the tensor corresponds to a directed edge, with the first row listing source nodes and the second row listing targets. For instance, the edge tensor ([[0], [9]]) from Figure~\ref{fig:cfg_dot} encodes a control-flow transition from node 0 (the opcode sequence) to node 9 (the exit block). This structure explicitly preserves the branching logic (e.g., jumps or loops) of the original smart contract.

\subsection{Data preprocessing}
\label{subsec:stage2}

Before feeding Solidity code into the model, we preprocess the source code to eliminate irrelevant information and enhance pattern recognition. This involves removing pragma statements, comments, non-ASCII characters, blank lines, and stopwords (e.g., ``function'', ``contract'') to focus on the code's functional elements. Additionally, we generalize function and variable names (e.g., ``FUN1'', ``VAR1'') to reduce noise from specific naming conventions, allowing the model to focus on structural and logical patterns rather than specific identifiers. These steps help streamline the input, reduce unnecessary complexity, and improve the model's ability to recognize key patterns and behaviors in the code.

Opcode is extracted from the Solidity compiler (Solc). During the opcode extraction process, the Solidity version of the contract is detected. If no version is found, Solc version 0.4.26 is used by default, as we have already inspected that all contracts in the dataset are written in the 0.4.x version. Next, any hexadecimal numbers (e.g., \texttt{0x40}) are removed, and the opcode is stripped of decimal characters and replaced with the original opcode. For example: \texttt{PUSH1 0x40} $\to$ \texttt{PUSH}, \texttt{DUP16} $\to$ \texttt{DUP}. Finally, the opcode is stripped of suffixes, as shown in Table~\ref{tab:opcode_mapping}.

\begin{table}[!ht]
    \centering
    \small
    \caption{Original and replaced opcode.}
    \begin{tabular}{ll}
        \hline
        \textbf{Original opcode} & \textbf{Replaced opcode} \\ \hline
        PUSH1 - PUSH32 & PUSH \\ \hline
        SWAP1 - SWAP16 & SWAP \\ \hline
        LOG1 - LOG4    & LOG  \\ \hline
        DUP1 - DUP16   & DUP  \\ \hline
    \end{tabular}
    \label{tab:opcode_mapping}
    \vspace{-2ex}
\end{table}

\subsection{Feature extraction}
\label{subsec:stage3}
\subsubsection{Enhancing CodeBERT with sliding window mechanism}
\begin{figure}[t]
    \centering
    \includegraphics[width=0.48\textwidth]{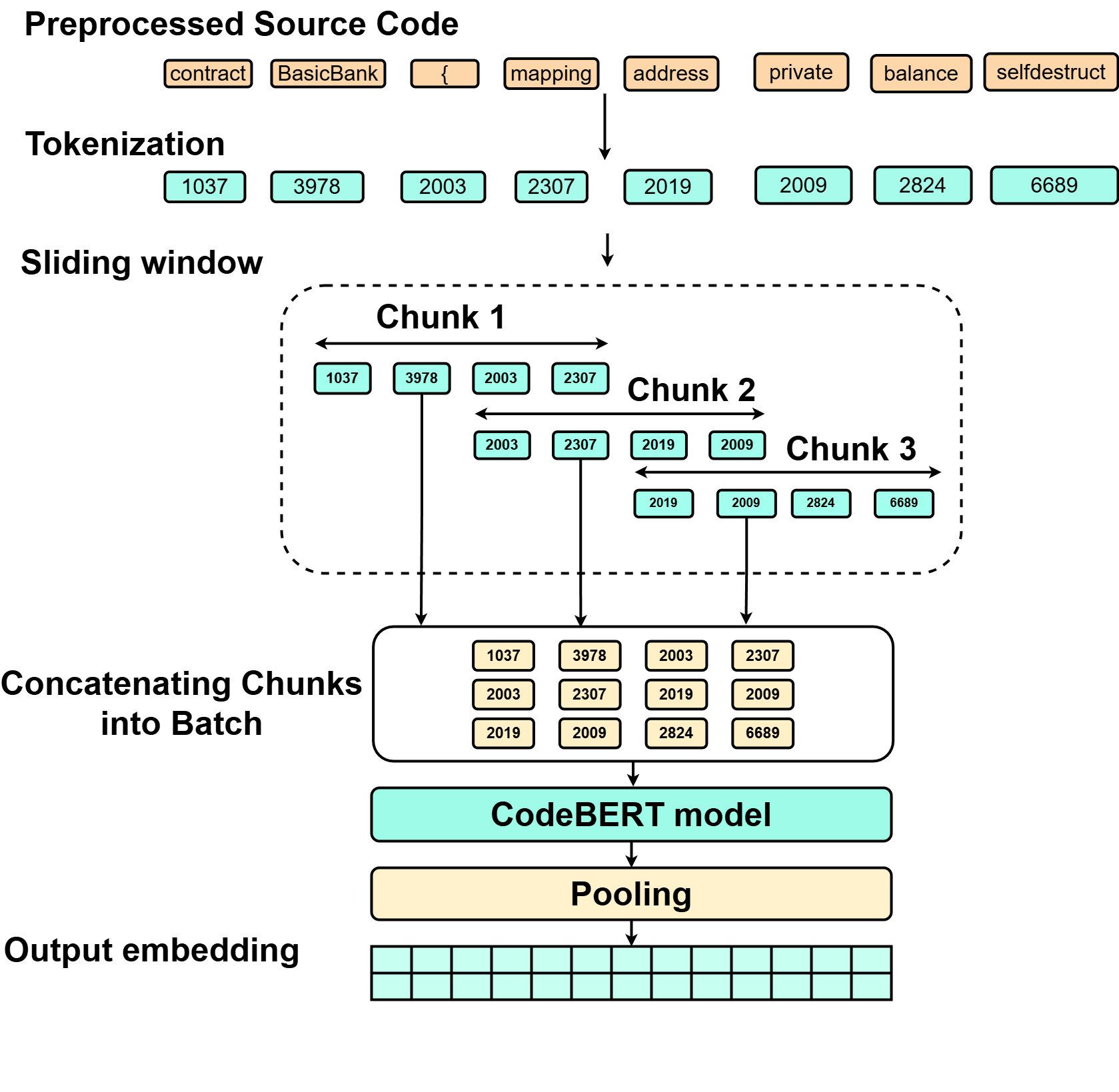}
    \caption{Sliding window processing pipeline.}
    \label{fig:sliding_codebert}
    \vspace{-2ex}
\end{figure}

CodeBERT \cite{feng2020codebert}, adapted from BERT \cite{bert} for programming languages, excels at bidirectional context understanding in code. We enhance it with a sliding window mechanism, as shown in \textbf{Figure \ref{fig:sliding_codebert}}, to process long sequences. The window (size 510 tokens for BERT's 512 limit) moves with a configurable stride - smaller strides preserve context through overlap, while incomplete final chunks are padded. The processing begins with WordPiece tokenization using CodeBERT's vocabulary, followed by sliding window segmentation where special tokens ([CLS]/[SEP]) are handled appropriately. The resulting chunks are then concatenated into batches for parallel processing through CodeBERT. After obtaining the results from all chunks, we apply max-pooling to extract the final embedding.

\subsubsection{xLSTM for opcode feature extraction}
The opcode feature extraction process effectively captures both local and global dependencies in the opcode sequences of smart contracts. We advocate for an xLSTM network, which captures long-range dependencies by maintaining context over time. The xLSTM component is particularly effective in modeling sequential relationships over larger spans, such as the influence of an earlier opcode on a later one in the contract execution flow.
The xLSTM architecture extends the traditional LSTM by enhancing its ability to retain information over longer sequences, which is crucial for smart contract analysis, where the relationship between distant opcodes can indicate specific vulnerability patterns. 

\begin{figure}[t]
\centering
\includegraphics[width=0.4\textwidth]{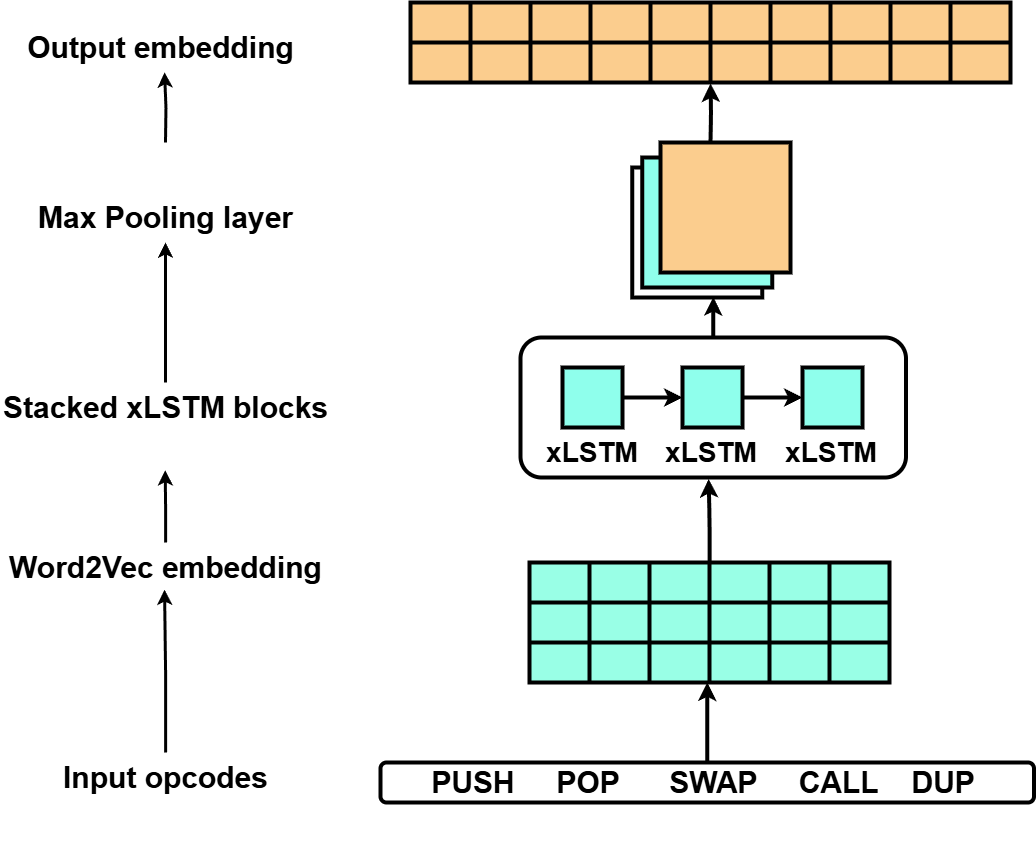}
\caption{Example of opcode feature extraction.}
\label{fig}
\vspace{-2ex}
\end{figure}

As shown in \textbf{Figure \ref{fig}}, the opcode sequences are first embedded using Word2Vec embeddings, which map each opcode to a dense vector representation, capturing semantic similarities between the opcodes. These embeddings are then fed into an xLSTM network for learning their representation.

\subsubsection{Graph attention neural network}

\begin{figure}[t]
    \centering
    \includegraphics[width=0.4\textwidth]{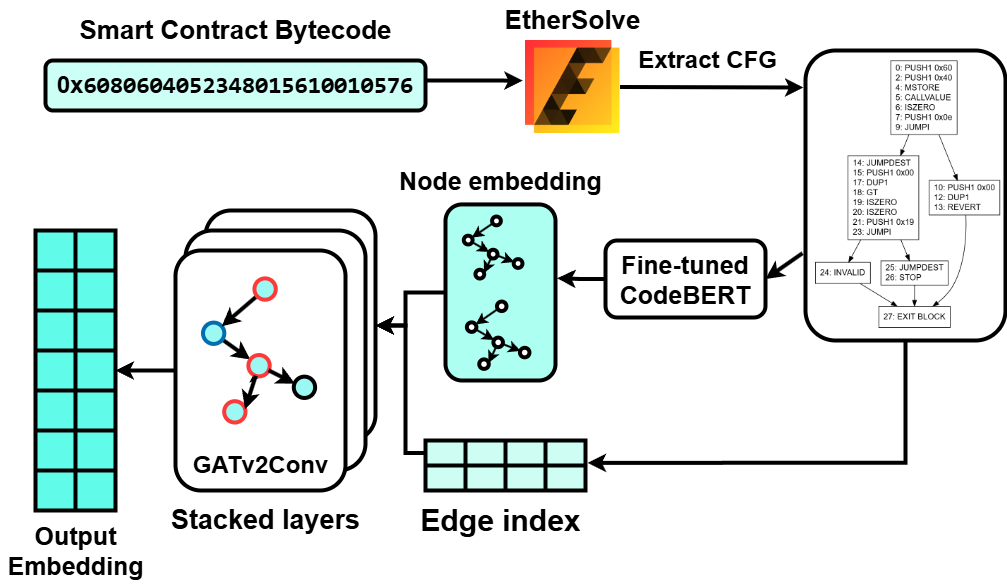} 
    \caption{Example of the bytecode transformation pipeline.}
    \label{fig:gat_flow}
    \vspace{-2ex}
\end{figure}

In our proposed framework, bytecode is processed using the EtherSolve \cite{contro2021ethersolve} tool to construct a CFG, as illustrated in \textbf{Figure \ref{fig:gat_flow}}. The CFG serves as a structured representation of the bytecode’s execution paths, detailing each operation, with nodes symbolizing instructions such as PUSH, JUMP, and CALLVALUE. These nodes, representing key instructions within the CFG, are subsequently embedded using a CodeBERT model fine-tuned on opcodes. The CodeBERT captures the semantic nuances of each operation, producing rich embeddings that encode operational context. The embedded nodes and the edge index are fed into a GATv2Conv model configured with stacked layers. This architecture enables the network to learn intricate relationships among nodes, effectively modeling the underlying structure of the graph. The resultant embedding from the GAT layer is then preserved for further concatenation steps, facilitating subsequent analyses and supporting classification tasks for vulnerability detection.

\subsection{Hierarchical cross-modal fusion}
\label{subsec:stage4}

We present a hierarchical fusion approach for integrating multiple modalities in vulnerability detection. Our approach comprises three distinct levels: intra-modal refinement through self-attention, pairwise cross-modal attention, and adaptive weighting. This architecture systematically combines information across modalities while managing computational complexity. Given modality embeddings $\mathbf{M}_1, \mathbf{M}_2, \dots, \mathbf{M}_N$, our fusion process operates as follows:

\subsubsection{Low-level fusion} 
At this level, we choose self-attention as the first fusion step due to its ability to capture fine-grained intra-modal dependencies within each modality, such as source code, opcode, and CFG. It enables each element to attend to context-relevant features regardless of position, which is crucial for structured and non-sequential data. Alternative fusion techniques like late fusion do not perform this internal refinement and are more appropriate for combining already processed representations at a higher level. For each modality $\mathbf{M}_i$, we compute:
\begin{equation}
    \mathbf{Z}_i = \text{MultiHeadAttention}(\mathbf{M}_i, \mathbf{M}_i, \mathbf{M}_i)
\end{equation}
The multi-head attention operation projects the input into query, key, and value spaces using learnable matrices:
\begin{equation}
    \mathbf{Q}_i = \mathbf{W}_Q \mathbf{M}_i, \quad
    \mathbf{K}_i = \mathbf{W}_K \mathbf{M}_i, \quad
    \mathbf{V}_i = \mathbf{W}_V \mathbf{M}_i
\end{equation}
where $\mathbf{W}_Q, \mathbf{W}_K, \mathbf{W}_V \in \mathbb{R}^{d \times d}$ are learnable weight matrices. Multiple attention heads operate in parallel, with each head $h$ computing attention scores as:
\begin{equation}
    \mathbf{A}_i^h = \text{softmax} \left( \frac{\mathbf{Q}_i^h (\mathbf{K}_i^h)^{\top}}{\sqrt{d_h}} \right)
\end{equation}
where $d_h = d / H$ is the dimension of head $h$ and $H$ is the total number of heads. The final self-attended representation is obtained by aggregating the outputs of all heads
\begin{equation}
    \mathbf{Z}_i = \mathbf{W}_O \text{Concat}(\mathbf{A}_i^1\mathbf{V}_i^1, \dots, \mathbf{A}_i^H\mathbf{V}_i^H)
\end{equation}
where $\mathbf{W}_O \in \mathbb{R}^{d \times d}$ is a learnable output projection matrix.

\subsubsection{Mid-level fusion} 
Following intra-modal refinement, we employ pairwise cross-attention to capture inter-modal interactions. This mechanism allows one modality to selectively attend to relevant features in another, enhancing complementary information exchange. Such targeted alignment is more effective than simpler fusion methods like concatenation at this stage. For each pair $(i,j)$ where $i \neq j$, we compute bidirectional cross-attention:
\begin{equation}
\begin{aligned}
\mathbf{C}_{i \rightarrow j} &= \text{PairWiseCrossAttention}(\mathbf{Z}_i, \mathbf{Z}_j, \mathbf{Z}_j) \\
\mathbf{C}_{j \rightarrow i} &= \text{PairWiseCrossAttention}(\mathbf{Z}_j, \mathbf{Z}_i, \mathbf{Z}_i)
\end{aligned}
\end{equation}
where $\mathbf{C}_{i \rightarrow j}$ represents the cross-attended features from modality $i$ attending to modality $j$, and vice versa for $\mathbf{C}_{j \rightarrow i}$. Similar to self-attention, the cross-attention mechanism computes:
\begin{equation}
    \mathbf{Q}_i = \mathbf{W}_Q^c \mathbf{Z}_i, \quad
    \mathbf{K}_j = \mathbf{W}_K^c \mathbf{Z}_j, \quad
    \mathbf{V}_j = \mathbf{W}_V^c \mathbf{Z}_j
\end{equation}
with the cross-attention scores for each head and final representation computed analogously to the self-attention mechanism. For each modality $i$, we aggregate the pairwise cross-attention outputs with the previous self-attention output as follows:
\begin{equation}
    \mathbf{H}_i = \mathbf{Z}_i + \frac{1}{N-1} \sum_{j \neq i} \mathbf{C}_{i \rightarrow j}
\end{equation}

This formulation ensures balanced information integration across modalities. Here, $N$ represents the total number of modalities. The normalization factor $\frac{1}{N - 1}$ prevents cross-attention features from dominating self-attention features as we scale to more modalities.

\subsubsection{High-level fusion} 
To complete the fusion process, we apply an adaptive weighting mechanism to combine the enhanced modality representations. Learnable weights \(\alpha_i\) enable the model to dynamically emphasize the most relevant modalities, facilitating flexible and task-driven integration after intra-modal and inter-modal dependencies have been captured. These weights are computed as:

\begin{equation}
    \alpha_i = \frac{\exp(w_i)}{\sum_{j=1}^{N} \exp(w_j)}
\end{equation}
where $w_i$ are trainable scalar parameters initialized equally across all modalities. In the next step, each modality representation is projected into a common semantic space through a non-linear transformation:
\begin{equation}
    \mathbf{H}_i' = \text{ReLU}(\mathbf{W}_i^f \mathbf{H}_i + \mathbf{b}_i^f)
\end{equation}
Here, $\mathbf{W}_i^f \in \mathbb{R}^{d \times d}$ and $\mathbf{b}_i^f \in \mathbb{R}^{d}$ are modality-specific learnable parameters that transform each enhanced representation $\mathbf{H}_i$ into a common feature space. Finally, the projected representations are combined using the learned adaptive weights:
\begin{equation}
    \mathbf{F} = \sum_{i=1}^{N} \alpha_i \mathbf{H}_i'
\end{equation}

This adaptive weighting mechanism prevents adversaries from targeting a single modality for obfuscation. By learning to distribute attention across multiple complementary modalities, the model maintains robustness even when vulnerabilities are deliberately concealed in one representation. 

\subsection{Classification}
\label{subsec:stage5}

After generating the fused representation $\mathbf{F}$ that integrates features across modalities, we employ a multi-label classification approach to identify potential vulnerabilities in smart contracts. Our framework is designed to detect four critical vulnerability categories simultaneously, including arithmetic vulnerabilities, low-level calls, time manipulation, and denial of service. Given the fused representation $\mathbf{F} \in \mathbb{R}^{d}$, we employ an MLP with a sigmoid activation function in the final layer:
\begin{equation}
    \mathbf{p} = \sigma(\mathbf{W}_2 \cdot \text{ReLU}(\mathbf{W}_1 \mathbf{F} + \mathbf{b}_1) + \mathbf{b}_2)
\end{equation}
where $\mathbf{W}_1 \in \mathbb{R}^{h \times d}$, $\mathbf{W}_2 \in \mathbb{R}^{4 \times h}$, $\mathbf{b}_1 \in \mathbb{R}^{h}$, and $\mathbf{b}_2 \in \mathbb{R}^{4}$ are learnable parameters, $h$ is the hidden dimension, and $\sigma$ represents the sigmoid activation function. The output $\mathbf{p} = [p_1, p_2, p_3, p_4]$ contains probability scores for each vulnerability type. For training, we employ the binary cross-entropy loss:
\begin{equation}
    \mathcal{L} = -\frac{1}{4}\sum_{i=1}^{4}[y_i\log(p_i) + (1-y_i)\log(1-p_i)]
\end{equation}
where $y_i \in \{0,1\}$ is the ground truth label for the $i$-th vulnerability type. During inference, we apply a threshold $\tau$ (set to 0.5 in our experiments):

\begin{equation}
    \hat{y}_i = 
    \begin{cases}
        1, & \text{if } p_i \geq \tau \\
        0, & \text{otherwise}
    \end{cases}
\end{equation}
This multi-label approach allows our model to detect multiple co-occurring vulnerabilities in a smart contract.

\section{Experiments and Analysis}\label{sec:experiments}

\subsection{Research questions}
To evaluate the effectiveness of the above research method, we propose the following research questions:

\begin{itemize}
    \item \textbf{RQ1:} How does the multi-modal model perform in smart contract vulnerability detection?
    \item \textbf{RQ2:} How robust is the proposed multi-modal model to obfuscated smart contracts?
    \item \textbf{RQ3:} How efficient is the multi-modal model in detecting vulnerabilities in terms of training and inference time?
\end{itemize}

\subsection{Experimental setup}
We evaluated ContractShield using seven models across three categories: unimodal, bimodal, and multimodal approaches. The models and fusion strategies are as follows:

\begin{itemize}
    \item \textbf{Unimodal:} M1 (CodeBERT on SC), M2 (xLSTM on OP), M3 (GATv2 on CFG)
    \item \textbf{Bimodal:} M1-M2 (SC + OP), M1-M3 (SC + CFG), M2-M3 (OP + CFG)
    \item \textbf{Multimodal:} ContractShield (M1-M2-M3): SC + OP + CFG
\end{itemize}

All experiments were conducted on an Intel Xeon CPU @ 2.20GHz (4 cores), NVIDIA Tesla P100, and 32GB RAM, using PyTorch and PyTorch Geometric. The hyperparameters are described in Table \ref{tab:hyperparams}.

\begin{table}[!ht]
\small
\centering
\caption{Experimental settings and hyperparameters.}
\label{tab:hyperparams}
\begin{tabular}{ll}
\hline
\textbf{Model} & \textbf{Hyperparameters} \\
\hline
\textbf{M1} & \begin{tabular}[c]{@{}l@{}}LR = $2 \times 10^{-5}$, batch = 2, \\ window = 9690 tokens, stride = 510\end{tabular} \\
\textbf{M2} & \begin{tabular}[c]{@{}l@{}}LR = $1 \times 10^{-3}$, batch = 64, \\ 2 xLSTM blocks, 512-dim embeddings\end{tabular} \\
\textbf{M3} & \begin{tabular}[c]{@{}l@{}}LR = $1 \times 10^{-4}$, batch = 64, \\ 3 GATv2 layers (128-dim, 8 heads)\end{tabular} \\
\textbf{Bimodal} & \begin{tabular}[c]{@{}l@{}}LR = $5 \times 10^{-4}$ to $1 \times 10^{-4}$, batch = 32–64\end{tabular} \\
\textbf{ContractShield} & \begin{tabular}[c]{@{}l@{}}LR = $2 \times 10^{-5}$, batch = 8, uses M1’s \\CodeBERT,  M2’s xLSTM, M3’s GATv2\end{tabular} \\
\hline
\end{tabular}
\end{table}

\subsection{Datasets and evaluation metrics}
\label{exp:datasets}

\subsubsection{Datasets} 
\label{sec:datasets}

\begin{table*}[!ht]
    \centering
    \caption{Vulnerability distribution across experimental datasets according to DASP Top 10 categories.}
    \label{table:combined_vulnerability_distribution}
    \small
    \begin{tabular}{lrrrrrr}
        \hline
        \multirow{2}{*}{\textbf{DASP Top 10 Category}} & 
        \multicolumn{2}{c}{\textbf{SoliAudit-SmartBugs}} & 
        \multicolumn{2}{c}{\textbf{CGT Weakness}} & 
        \multicolumn{2}{c}{\textbf{DAppScan}} \\
        \cline{2-7}
        & \textbf{Pos.} & \textbf{Neg.} & \textbf{Pos.} & \textbf{Neg.} & \textbf{Pos.} & \textbf{Neg.} \\
        \hline
        Reentrancy & 9,337 & 14,002 & 1,175 & 814 & 8 & 116 \\
        Arithmetic & 21,366 & 1,973  & 681  & 1,308 & 20 & 104 \\
        Unchecked Low-Level Calls & 11,702 & 11,637 & 1,240 & 749 & 3 & 121 \\
        Denial of Service & 8,070 & 15,269 & 1,188 & 801 & 10 & 114 \\
        Time Manipulation & 7,815 & 15,524 & 110 & 1,879 & 7 & 117 \\
        \hline
        \textbf{Total Samples} & \multicolumn{2}{c}{\textbf{23,339}} & 
                                 \multicolumn{2}{c}{\textbf{1,989}} & 
                                 \multicolumn{2}{c}{\textbf{124}} \\
        \hline
    \end{tabular}
\end{table*}

The evaluation focuses on five vulnerability types aligned with the DASP Top-10 categories: \emph{Reentrancy}, \emph{Unchecked Low-Level Calls}, \emph{Denial of Service}, \emph{Time Manipulation}, and \emph{Arithmetic}. All datasets used in the study exhibit a multi-label structure, as a single smart contract may contain multiple vulnerability categories. To enable comprehensive training and evaluation, three corpora were prepared following a unified preprocessing pipeline that includes opcode extraction, bytecode normalization, and CFG construction.

\begin{itemize}
    \item \textbf{SoliAudit \cite{liao2019soliaudit} + SmartBug-Wilds \cite{ferreira2020smartbugs}.} This combined corpus includes 17,142 contracts from SoliAudit and 47,398 contracts from SmartBugs-Wilds. After preprocessing and removing samples that failed to compile or contained unresolved dependencies, 23,339 valid contracts were obtained. A total of 18,671 contracts were allocated to the training set using an 80\% stratified split, and the remaining 4,668 contracts formed \textbf{Test Set A}.
    
    \item \textbf{CGT Weakness \cite{di2023consolidation}.} The original dataset contains 3,103 manually validated contracts drawn from multiple public sources. Following preprocessing and CFG conversion, 1,989 contracts were successfully retained. This corpus serves exclusively as \textbf{Test Set B} for evaluating generalization to an independent distribution.

    \item \textbf{DAppScan \cite{zheng2024dappscan}.} This real-world dataset was built from 1,199 professional audit reports covering 682 decentralized applications. The DAPPSCAN-SOURCE subset consists of 947 annotated Solidity source files. Due to the complex dependency structures commonly found in large DApp projects, only 124 contracts were successfully compiled, processed, and transformed into CFGs. These contracts constitute \textbf{Test Set C}, which enables evaluation under realistic software-engineering and deployment conditions.
\end{itemize}

\textbf{Table \ref{table:combined_vulnerability_distribution}} summarizes the distribution of vulnerability labels across all datasets. For each DASP category, the table reports the number of \emph{positive} (Pos.) samples in which a vulnerability is present and the number of \emph{negative} (Neg.) samples in which it is absent. These statistics guided the choice of target labels and informed the stratified data-splitting strategy used throughout the study.

\paragraph{Training and testing setup}
All models are trained on the 18,671-contract training split of the combined SoliAudit and SmartBugs-Wilds corpus. Evaluation is performed on three test sets:

\begin{table}[!ht]
    \small
    \caption{Performance metrics.}
    \centering
    \renewcommand{\arraystretch}{1.05}
    \begin{tabular}{p{0.23\columnwidth} p{0.64\columnwidth}}
    \hline
    \textbf{Metric} & \textbf{Description} \\
    \hline
    True Positive (TP) & Labels correctly predicted as present \\
    True Negative (TN) & Labels correctly predicted as absent \\
    False Positive (FP) & Labels incorrectly predicted as present \\
    False Negative (FN) & Labels incorrectly predicted as absent \\
    Hamming Score (HS) & Fraction of labels correctly predicted \\
    Precision (Pre) & TP divided by TP plus FP \\
    Recall (Rec) & TP divided by TP plus FN \\
    F1-score (F1) & Harmonic mean of Precision and Recall \\
    HS Degradation & Difference between baseline HS and obfuscated HS \\
    \hline
    \end{tabular}
    \label{tab:metrics}
\end{table}

All approaches are evaluated using standard classification metrics including Precision (Pre), Recall (Rec), F1-score (F1), and Hamming Score (HS). \textbf{Table \ref{tab:metrics}} summarizes the definition of each metric. The reported results are obtained by averaging metric values across all samples and their associated labels.

Because multi-label prediction requires evaluation at the level of individual labels, HS is used as a primary metric. This measure quantifies the proportion of correctly predicted labels, treating each label independently, and is therefore well-suited for scenarios where class distributions may be imbalanced or where the presence of multiple simultaneous labels must be captured.

For a test set with $n$ samples and $L$ labels, let $y_{i,j}$ denote the ground-truth value and $\hat{y}_{i,j}$ the predicted value of the $j$th label for the $i$th instance. The HS is computed as:
\begin{equation}
    \text{Hamming Score} = \frac{1}{n L} \sum_{i=1}^{n} \sum_{j=1}^{L} \mathbb{1}(y_{i,j} = \hat{y}_{i,j})
\end{equation}
where $\mathbb{1}(\cdot)$ is the indicator function equal to 1 when the values match and 0 otherwise. HS ranges from 0 to 1, with higher scores indicating better classification performance.

To assess the influence of obfuscation techniques on model robustness, Hamming Score Degradation is additionally reported. This metric measures the reduction in predictive accuracy under obfuscation relative to the non-obfuscated baseline. For a given model, let $\text{HS}_{\text{base}}$ be the HS on clean data and $\text{HS}_{\text{obf}}$ the HS obtained under a specific obfuscation setting. The degradation is defined as:
\begin{equation}
    \text{HS Degradation} = \text{HS}_{\text{base}} - \text{HS}_{\text{obf}}
\end{equation}

Higher degradation values reflect a stronger negative impact from obfuscation, whereas lower values indicate that the model is more resilient to perturbations introduced by transformations in the source code or bytecode.

\subsection{Robustness evaluation against code obfuscation}
\label{sec:code_obfuscation_evaluation}
To evaluate model robustness in realistic adversarial conditions, we apply code obfuscation techniques to all evaluation sets defined in \textbf{Section \ref{sec:datasets}}. For each test set (A, B, and C), both clean and obfuscated variants are constructed, enabling a direct comparison of performance under realistic adversarial transformations while avoiding any training-time contamination.

\begin{figure*}[!ht]
\centering 
\includegraphics[width=0.9\textwidth]{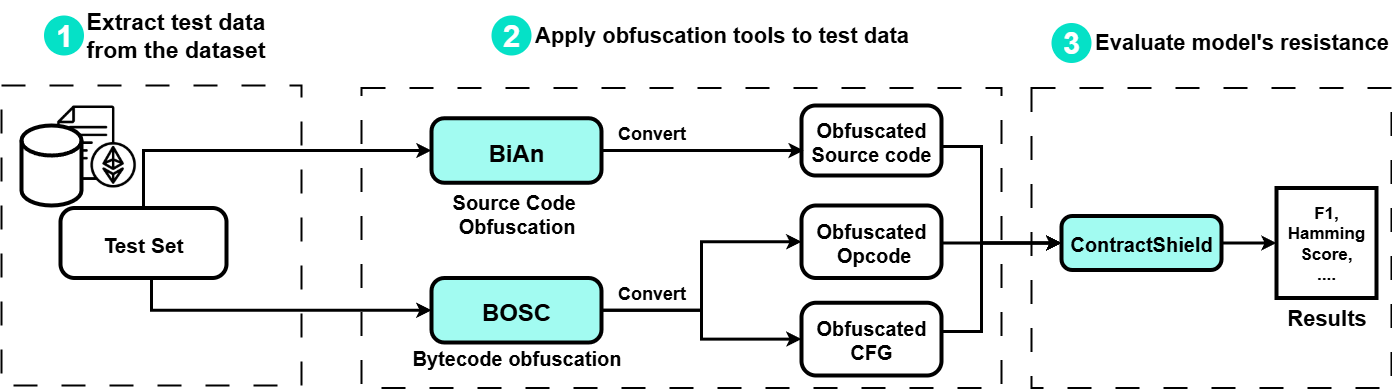} 
\caption{Three-step process for evaluating ContractShield's resistance to obfuscation techniques.} 
\label{fig:obfuscation_process} 
\end{figure*}

Obfuscation is applied at two abstraction levels, specifically the source code level and the bytecode level, enabling consistent robustness evaluation across all modalities. Corresponding CFGs and opcode sequences are also generated from the obfuscated bytecode to ensure full coverage of structural and sequential representations. The overall workflow is illustrated in Figure \ref{fig:obfuscation_process}.

\subsubsection{Source code obfuscation}
Source code obfuscation is performed using the BiAn tool \cite{zhang2023bian}, which has been shown to markedly increase contract complexity and reduce the effectiveness of decompilers and static analysis tools. It applies a structured three-step process targeting Solidity contracts:

\begin{itemize}
    \item \textbf{Control flow obfuscation:} This technique complicates the program's execution flow through two primary methods. First, \textit{opaque predicate insertion} introduces conditional statements with predetermined truth values using CPM chaotic mapping, creating false execution paths that confuse static analysis tools. Second, \textit{control flow graph flattening} restructures the program by breaking basic blocks and reconstructing them with jump-based execution, effectively destroying the original logical structure and making reverse engineering significantly more challenging.
    
    \item \textbf{Data flow obfuscation:} This component modifies data representations without altering program logic through five transformation strategies: (1) \textit{Local-to-global variable conversion} transforms local variables into global state variables, increasing program complexity; (2) \textit{Static-to-dynamic data transformation} converts constants into dynamic array-based lookups; (3) \textit{Integer constant expansion} replaces simple constants with complex arithmetic expressions; (4) \textit{Boolean variable splitting} decomposes boolean values using logical operators; and (5) \textit{Scalar-to-vector transformation} reorganizes individual variables into structured data types.
    
    \item \textbf{Layout obfuscation:} This final step removes human-readable information by: (1) \textit{Comment deletion} to eliminate explanatory text; (2) \textit{Layout scrambling} to disrupt code formatting; and (3) \textit{Variable name replacement} using SHA-1 hashing to generate cryptographic identifiers, making the code significantly less readable while maintaining functional equivalence.
\end{itemize}

\begin{figure}[!ht]
\centering
\begin{subfigure}{0.52\textwidth}
\centering
\includegraphics[width=\textwidth]{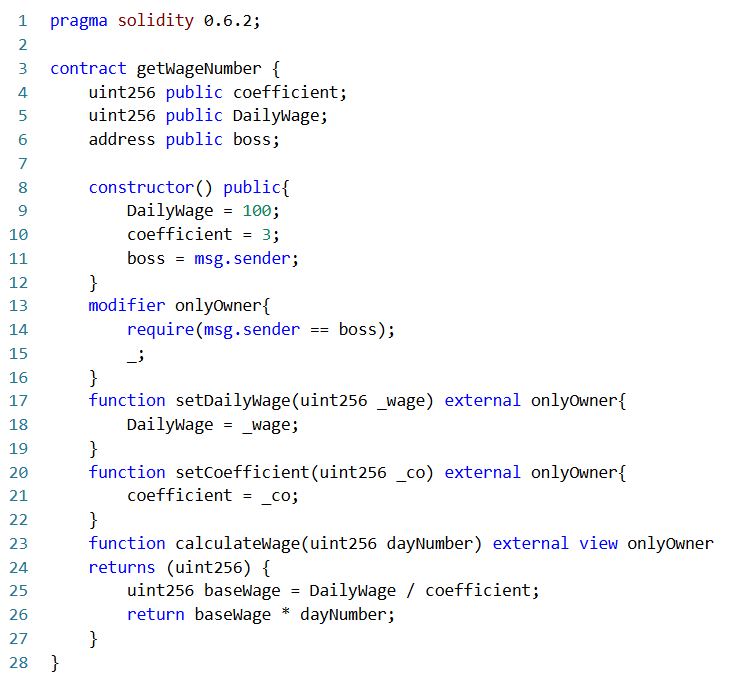}
\caption{Original smart contract.}
\label{fig:original}
\end{subfigure}\\
\begin{subfigure}{0.48\textwidth}
\centering
\includegraphics[width=\textwidth]{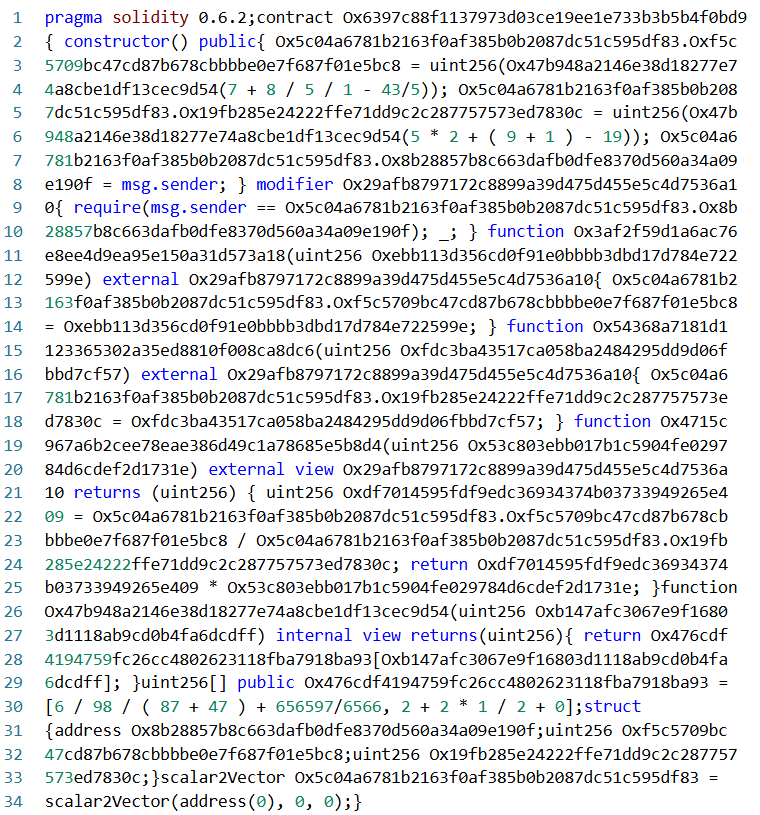}
\caption{Smart contract after obfuscation.}
\label{fig:obfuscated}
\end{subfigure}
\caption{Examples of original and obfuscated contracts using BiAn.}
\label{fig:source_obfuscation}
\end{figure}

A sample contract before and after BiAn obfuscation is shown in \textbf{Figure \ref{fig:source_obfuscation}}. BiAn applies several techniques that significantly increase code complexity. First, all meaningful identifiers such as contract names (\texttt{getWageNumber}), functions (\texttt{calculat-\\eWage}, \texttt{setDailyWage}), variables (\texttt{coefficient}, \texttt{DailyWage}), and modifiers (\texttt{onlyOwner}) are replaced with long hexadecimal strings that resemble SHA-1 outputs (for example \texttt{Ox6397c88f1\\137973d...} or \texttt{Ox4715c967a6b2cee78...}). Second, integer constants are rewritten as arithmetic expressions. For example, \texttt{100} becomes \texttt{5 * 2 + (9 + 1) - 19}, and \texttt{3} becomes \texttt{7 + 8 / 5 / 1 - 43/5}. Third, state variables are moved into a struct and accessed through a global struct instance, turning variables like \texttt{coefficient} into fields such as \texttt{Ox5c04a6781b2163f0af\\385b0b2087dc51c595df83.Oxf5c5709bc47cd87b67856}. \\Fourth, literal values are relocated into arrays (for example \texttt{[6/98/(87+47) + 656597/6566, 2 + 2 * 1 / 2 + 0]}) \\and retrieved via helper functions like \texttt{Ox47b948a2146e...()}. Finally, all formatting is removed, including whitespace, indentation, and comments, resulting in dense continuous text. As shown in \textbf{Figure \ref{fig:obfuscated}}, the readable 39-line contract is compressed into a 34-line cryptic artifact that lacks clear semantic structure and is difficult for both humans and automated tools to analyze.

\subsubsection{Bytecode obfuscation}
For bytecode-level obfuscation, we employ the BOSC (Bytecode Obfuscation for Smart Contracts) tool developed by Yu et al. \cite{yu2022bytecode}, which transforms runtime bytecode through a three-phase pipeline:

\begin{itemize}
    \item \textbf{Preprocessing:} BOSC extracts runtime bytecode from the complete bytecode file, which contains deployment, runtime, and auxiliary data components. Since decompilation tools typically analyze only the runtime bytecode that contains the core program logic, BOSC focuses its obfuscation efforts on this critical component.

    \item \textbf{Bytecode obfuscation techniques:} BOSC applies four complementary transformations: (1) \textit{Incomplete instruction obfuscation} inserts opcode-only partial instructions to disrupt decompiler parsing; (2) \textit{False branch obfuscation} converts unconditional jumps into conditional ones with unreachable branches, creating infinite loops that exhaust decompiler resources; (3) \textit{Instruction sequence rearrangement} reorders independent instructions to obscure developer logic patterns; and (4) \textit{Flower instruction obfuscation} injects meaningless "junk" instructions that complicate program flow analysis without affecting execution.

    \item \textbf{Recovery:} After obfuscating the runtime bytecode, BOSC reconstructs the full deployable bytecode by reintegrating the previously removed deployment and auxiliary sections.
\end{itemize}

After obfuscation, the resulting bytecode is used to derive both CFGs and opcode sequences, enabling our multimodal evaluation framework to assess model robustness across multiple representations: source code (via BiAn), bytecode (via BOSC), CFGs (derived from obfuscated bytecode), and opcodes (extracted from obfuscated bytecode). This ensures a comprehensive evaluation of resistance to obfuscation at different abstraction layers of smart contract analysis.

\subsection{Experimental results}
\label{experiment_results}
\subsubsection{Results on non-obfuscated data (RQ1)}

Before testing robustness under code obfuscation, we first examine ContractShield’s detection accuracy on three non-obfuscated benchmark datasets: \textbf{SoliAudit+SmartBug}, \textbf{CGTWeakness}, and \textbf{DAppScan}, corresponding respectively to Test set A, B, and C. The Test set A dataset serves as the major reference benchmark due to its balanced class distribution.

\textbf{Table~\ref{tab:performance_metrics_modalities_non_obf}} presents a detailed comparison of all seven model configurations on Test set A. The unimodal results reveal clear differences in the predictive capacity of individual modalities. M1 (CodeBERT on source code) achieves the strongest unimodal performance with the lowest HL (0.068), the highest HS (89.01\%), and an F1-score of 91\%. These results reflect the expressiveness of high-level source code, which preserves semantic structure, data-flow behavior, and vulnerability-related patterns that transformer models can effectively exploit. M2 (xLSTM on opcode) achieves an F1-score of 87\%, together with higher error rates and lower label-wise accuracy compared to M1. This performance gap is expected, as opcode sequences capture low-level execution behavior but lack the semantic and syntactic cues present in source code, making the signals more fragmented and less informative for precise vulnerability detection. M3 (GATv2 on CFG) yields the weakest unimodal performance, reflected in its higher HL, an HS of just over 80\%, and the lowest F1-score with 81.32\%. While control-flow topology captures structural program logic, it alone cannot represent functional semantics such as data dependencies or contract-specific invariants, limiting its ability to support precise vulnerability classification.

The bimodal configurations show that combining two modalities yields moderate gains but still falls short of full cross-modal complementarity. The M1–M2 model, which fuses source code and opcode information, achieves an F1-score close to 90\%, but its performance remains slightly below M1, particularly in label-wise accuracy. This indicates that opcode sequences contribute additional low-level execution cues, although much of their information overlaps with what is already present in high-level source code. The M2–M3 combination exhibits a similar trend, achieving an F1-score of 88.58\% and improving both HS and HL compared to using opcode or CFG alone. Here, structural signals from the CFG help compensate for the limited semantics in opcode sequences, but the fusion still lacks the richer contextual information available from source code. The weakest bimodal pairing is M1–M3. Despite combining source semantics with structural topology, its performance remains lower than M1 alone. This is because CFGs provide only coarse control-flow skeletons that lack the fine-grained semantic cues necessary for precise vulnerability detection, so their contribution is limited when fused with source code features.

Our multimodal approach, ContractShield, which fuses all three types of features through hierarchical cross-modal fusion, achieves overall performance comparable to the best unimodal model (M1). The results indicate that source code remains the most influential modality, while the additional signals from opcode and CFG do not provide a noticeable improvement in detection performance on clean data. However, compared to the bimodal models, ContractShield demonstrates a more balanced performance across HL, HS, precision, and recall, indicating that the fusion effectively consolidates complementary patterns from all modalities. To better evaluate the contribution of opcode and CFG, RQ2 conducts experiments on obfuscated datasets, which will clarify their impact on model robustness.

\begin{table}[!ht]
    \centering
    \caption{Performance results comparison across different modalities on Test-set A.}
    \resizebox{0.5\textwidth}{!}{ 
        \begin{tabular}{lccccc} 
            \toprule
            \textbf{Model} & \textbf{HL} & \textbf{HS (\%)} & \textbf{F1 (\%)} & \textbf{Pre (\%)} & \textbf{Rec (\%)} \\
            \midrule
            M1 & 0.068 & 89.01 & 91.00 & 90.28 & 91.73 \\
            M2 & 0.134 & 86.58 & 87.00 & 88.12 & 85.90 \\
            M3 & 0.162 & 80.53 & 81.32 & 82.55 & 80.12 \\
            M1-M2 & 0.078 & 88.43 & 90.06 & 90.91 & 89.22 \\
            M1-M3 & 0.125 & 85.33 & 86.13 & 87.18 & 85.10 \\
            M2-M3 & 0.122 & 87.92 & 88.58 & 89.35 & 87.82 \\
            \textbf{ContractShield} & \textbf{0.058} & \textbf{89.16} & \textbf{91.47} & \textbf{91.12} & \textbf{91.83} \\
            \bottomrule
        \end{tabular}
    }
    \label{tab:performance_metrics_modalities_non_obf}
\end{table}

We compare our proposed method with three existing works, as reported in \textbf{Table \ref{tab:performance_metrics_method_non_obf}}. Qian et al. \cite{qian_auto_squen_mol} introduced a sequential approach for reentrancy detection, which operates on smart contract source code. Their approach applies a bidirectional LSTM with an attention mechanism (BLSTM-ATT) to precisely detect reentrancy bugs. Cheong et al. \cite{cheong2024gnn} also proposed a multi-label vulnerability detection model based on GNNs, where smart contract opcodes are preprocessed and transformed into control flow graphs that serve as input to the GNN for multi-label classification. Deng et al. \cite{deng2023smart} presented a multimodal deep learning framework for smart contract vulnerability detection, which incorporates source code semantics, opcode sequences, and control-flow structure. Their approach fuses these modalities using a concatenation-based decision fusion mechanism to perform multi-label classification.

For a fair comparison, all three baseline architectures were reconstructed based on the descriptions in their respective papers, and trained and evaluated on the same dataset detailed in \textbf{Section \ref{sec:datasets}}.

\begin{table}[!ht]
\centering
\caption{Performance comparison between ContractShield and other state-of-the-art methods across different datasets.}
\label{tab:performance_metrics_method_non_obf}
\resizebox{0.5\textwidth}{!}{%
\begin{tabular}{llcccc}
\toprule
\textbf{Model} & \textbf{Test set} & \textbf{HS(\%)} & \textbf{F1 (\%)} & \textbf{Pre (\%)} & \textbf{Rec (\%)} \\
\midrule

\multirow{3}{*}{Qian et al. \cite{qian_auto_squen_mol}}
& Test set A & 78.15 & 78.87 & 87.79 & 79.26 \\
& Test set B & 70.42 & 74.67 & 78.96 & 70.82 \\
& Test set C & 61.37 & 65.01 & 68.45 & 61.90 \\
\midrule

\multirow{3}{*}{Cheong et al. \cite{cheong2024gnn}}
& Test set A & 70.61 & 74.35 & 80.91 & 70.16 \\
& Test set B & 64.05 & 67.87 & 71.63 & 64.41 \\
& Test set C & 57.12 & 60.57 & 63.82 & 57.46 \\
\midrule

\multirow{3}{*}{Deng et al. \cite{deng2023smart}}
& Test set A & 81.60 & 84.85 & 85.46 & 84.12 \\
& Test set B & 75.27 & 77.30 & 78.05 & 76.58 \\
& Test set C & 67.18 & 69.11 & 69.84 & 68.41 \\
\midrule

\multirow{3}{*}{\textbf{ContractShield}}
& \textbf{Test set A} & \textbf{89.16} & \textbf{91.47} & \textbf{91.12} & \textbf{91.83} \\
& \textbf{Test set B} & \textbf{81.03} & \textbf{83.47} & \textbf{84.09} & \textbf{82.87} \\
& \textbf{Test set C} & \textbf{73.36} & \textbf{75.72} & \textbf{76.04} & \textbf{75.41} \\
\bottomrule
\end{tabular}}
\end{table}

The results in \textbf{Table \ref{tab:performance_metrics_method_non_obf}} show that ContractShield consistently delivers the strongest performance across all three test sets, outperforming all baseline methods by a clear margin. Compared with Qian et al. and Cheong et al., both of which rely on a single input modality, the improvements are substantial. On Test set A, for example, ContractShield achieves an F1-score of 91.47\% and an HS of 89.16\%, substantially outperforming both single-modality baselines. Qian’s opcode-based sequential model reaches only 78.87\% F1 and 78.15\% HS, while Cheong’s CFG-based GNN performs even lower with 74.35\% F1 and 70.61\% HS. Similar performance gaps appear on Test sets B and C, confirming that single-modality representations, such as opcode sequences or CFGs, capture only partial aspects of program behavior and cannot match the coverage provided by a more comprehensive multimodal representation.

Deng et al.’s multimodal framework narrows the gap but still falls notably short of ContractShield across all evaluation settings. Although their approach incorporates multiple modalities, the fusion mechanism is based on simple concatenation, which merely appends features from different sources into a single vector without modeling their semantic correspondence or structural alignment. This type of fusion yields a shallow joint representation that fails to capture deeper cross-modal interactions, allowing dominant modalities to overshadow weaker yet complementary ones \cite{sahu2021adaptive, upadhya2024vulnfusion}. As a result, the fused features fail to express the intermodal dependencies that are crucial for multi-label vulnerability detection. This limitation is reflected in Deng’s Test set A F1-score of 84.85\%, which remains well below the 91.47\% achieved by ContractShield.

\begin{table*}[!ht]
\small
\centering
\caption{Performance comparison across different modality combinations under obfuscation on Test set A.}
\label{tab:performance_metrics_modalities_obf}
\resizebox{0.85\textwidth}{!}{%
\begin{tabular}{ccccccc}
\hline
\textbf{Model} &
  \textbf{Obfuscation} &
  \textbf{HS (\%)} &
  \textbf{F1 (\%)} &
  \textbf{Precision (\%)} &
  \textbf{Recall (\%)} &
  \textbf{HS Degradation (\%)} \\ \hline
M1                     & BiAn               & 84.25          & 87.31          & 87.91          & 86.72          & 4.76          \\ \hline
M2                     & BOSC               & 82.10          & 84.48          & 87.01          & 82.10          & 4.48          \\ \hline
M3                     & BOSC               & 77.23          & 82.20          & 79.37          & 85.25          & 3.30          \\ \hline
\multirow{3}{*}{M1-M2} & BiAn               & 86.67          & 86.91          & 88.59          & 85.29          & 1.76          \\
                       & BOSC               & 84.10          & 85.41          & 85.90          & 84.92          & 4.33          \\
                       & BiAn+BOSC          & 83.25          & 83.55          & 84.01          & 83.11          & 5.18          \\ \hline
\multirow{3}{*}{M1-M3} & BiAn               & 83.85          & 83.74          & 86.76          & 80.92          & 1.48          \\
                       & BOSC               & 81.34          & 82.45          & 84.22          & 80.75          & 3.99          \\
                       & BiAn+BOSC          & 82.20          & 83.77          & 84.92          & 82.66          & 3.13          \\ \hline
M2-M3                  & BOSC               & 86.34          & 89.34          & 88.26          & 90.45          & 1.58          \\ \hline
\multirow{3}{*}{\textbf{ContractShield}} &
  \textbf{BiAn} &
  \textbf{86.97} &
  \textbf{87.94} &
  \textbf{89.00} &
  \textbf{86.91} &
  \textbf{2.19} \\
                       & \textbf{BOSC}      & \textbf{85.74} & \textbf{86.73} & \textbf{87.90} & \textbf{85.60} & \textbf{3.42} \\
                       & \textbf{BiAn+BOSC} & \textbf{87.94} & \textbf{88.65} & \textbf{89.50} & \textbf{87.82} & \textbf{1.22} \\ \hline
\end{tabular}}
\end{table*}

\subsubsection{Results on obfuscated data (RQ2)}

To evaluate the stability of different feature representations and to determine whether multimodal learning truly provides stronger resilience beyond clean-data accuracy, RQ2 evaluates model robustness under three obfuscation scenarios: source-code obfuscation using BiAn \cite{zhang2023bian}, bytecode-level obfuscation using BOSC \cite{yu2022bytecode}, and a combined setting (BiAn+BOSC), as detailed in \textbf{Section \ref{sec:code_obfuscation_evaluation}}. The performance across these settings is summarised in \textbf{Table \ref{tab:performance_metrics_modalities_obf}}. 

Across all three obfuscation settings, the overall impact on model performance remains modest, with HS degradation ranging from 1.22\% to 4.76\%, indicating that the models maintain strong stability even under adversarial perturbations. Under the BiAn scenario, the source-code unimodal model M1 exhibits the largest drop, yet the reduction remains moderate, as CodeBERT is inherently robust to lexical and syntactic variations. Under the BOSC setting, the impact shifts toward models that rely on bytecode features, where the opcode-based M2 and CFG-based M3 experience comparatively greater sensitivity.

When these feature representations are fused in bimodal configurations or fully fused within the multimodal ContractShield model, the impact of obfuscation is further reduced. ContractShield, in particular, demonstrates consistently high robustness, with HS decreasing only between 1.22\% and 3.42\%, while still maintaining stable performance levels (HS of 85.74\%–87.94\% and F1-scores of 86.73\%–87.94\%). These results highlight the complementary strengths of different feature modalities and confirm that multimodal fusion significantly enhances resilience against diverse obfuscation techniques.



\begin{table*}[!ht]
\centering
\caption{Performance comparison of methods under different obfuscation settings.}
\label{tab:performance_metrics_method_obf}
\small
\begin{tabular}{cccccccc}
\hline
\textbf{Model} &
  \textbf{Test set} &
  \textbf{Obfuscation} &
  \textbf{HS (\%)} &
  \textbf{F1 (\%)} &
  \textbf{Precision (\%)} &
  \textbf{Recall (\%)} &
  \textbf{HS Degradation} \\ \hline
\multirow{3}{*}{Qian et al. \cite{qian_auto_squen_mol}} &
  Test set A &
  Bian &
  75.21 &
  79.91 &
  84.23 &
  76.02 &
  2.94 \\ \cline{2-8} 
 &
  Test set B &
  Bian &
  67.14 &
  72.31 &
  77.03 &
  68.54 &
  3.28 \\ \cline{2-8} 
 &
  Test set C &
  Bian &
  59.88 &
  65.11 &
  71.22 &
  60.13 &
  4.13 \\ \hline
\multirow{3}{*}{Cheong et al. \cite{cheong2024gnn}} &
  Test set A &
  BOSC &
  66.15 &
  72.96 &
  78.51 &
  68.15 &
  4.46 \\ \cline{2-8} 
 &
  Test set B &
  BOSC &
  61.02 &
  67.23 &
  74.25 &
  61.89 &
  5.13 \\ \cline{2-8} 
 &
  Test set C &
  BOSC &
  54.82 &
  60.27 &
  67.00 &
  55.38 &
  5.79 \\ \hline
\multirow{9}{*}{Deng et al. \cite{deng2023smart}} &
  \multirow{3}{*}{Test set A} &
  BiAn &
  79.45 &
  81.30 &
  82.90 &
  79.77 &
  2.15 \\
 &
   &
  BOSC &
  76.12 &
  78.62 &
  80.45 &
  76.88 &
  5.48 \\
 &
   &
  BiAn + BOSC &
  77.55 &
  79.33 &
  81.33 &
  77.42 &
  4.05 \\ \cline{2-8} 
 &
  \multirow{3}{*}{Test set B} &
  BiAn &
  72.18 &
  74.41 &
  76.31 &
  72.55 &
  3.09 \\
 &
   &
  BOSC &
  68.75 &
  71.83 &
  73.21 &
  69.22 &
  6.25 \\
 &
   &
  BiAn + BOSC &
  70.84 &
  72.92 &
  74.83 &
  71.11 &
  4.43 \\ \cline{2-8} 
 &
  \multirow{3}{*}{Test set C} &
  BiAn &
  65.30 &
  67.40 &
  69.18 &
  65.67 &
  3.86 \\
 &
   &
  BOSC &
  61.15 &
  63.79 &
  65.48 &
  61.62 &
  6.03 \\
 &
   &
  BiAn + BOSC &
  63.23 &
  65.09 &
  67.33 &
  63.87 &
  4.55 \\ \hline
\multirow{9}{*}{\textbf{ContractShield}} &
  \multirow{3}{*}{\textbf{Test set A}} &
  \textbf{BiAn} &
  \textbf{87.02} &
  \textbf{87.94} &
  \textbf{89.00} &
  \textbf{86.91} &
  \textbf{2.14} \\
 &
   &
  \textbf{BOSC} &
  \textbf{85.74} &
  \textbf{86.73} &
  \textbf{87.90} &
  \textbf{85.60} &
  \textbf{3.42} \\
 &
   &
  \textbf{BiAn + BOSC} &
  \textbf{87.94} &
  \textbf{88.65} &
  \textbf{89.50} &
  \textbf{87.82} &
  \textbf{1.22} \\ \cline{2-8} 
 &
  \multirow{3}{*}{\textbf{Test set B}} &
  \textbf{BiAn} &
  \textbf{82.35} &
  \textbf{83.48} &
  \textbf{84.21} &
  \textbf{82.78} &
  \textbf{1.81} \\
 &
   &
  \textbf{BOSC} &
  \textbf{80.94} &
  \textbf{81.73} &
  \textbf{82.47} &
  \textbf{80.24} &
  \textbf{3.22} \\
 &
   &
  \textbf{BiAn + BOSC} &
  \textbf{82.76} &
  \textbf{83.88} &
  \textbf{84.61} &
  \textbf{83.18} &
  \textbf{1.64} \\ \cline{2-8} 
 &
  \multirow{3}{*}{\textbf{Test set C}} &
  \textbf{BiAn} &
  \textbf{75.27} &
  \textbf{76.48} &
  \textbf{77.00} &
  \textbf{75.98} &
  \textbf{1.75} \\
 &
   &
  \textbf{BOSC} &
  \textbf{73.96} &
  \textbf{75.04} &
  \textbf{75.74} &
  \textbf{74.24} &
  \textbf{3.03} \\
 &
   &
  \textbf{BiAn + BOSC} &
  \textbf{75.62} &
  \textbf{76.91} &
  \textbf{77.51} &
  \textbf{76.43} &
  \textbf{1.57} \\ \hline
\end{tabular}
\end{table*}


\subsubsection{Experiment for RQ3}

\begin{figure}[!ht]
  \centering
  \includegraphics[width=1.0\linewidth]{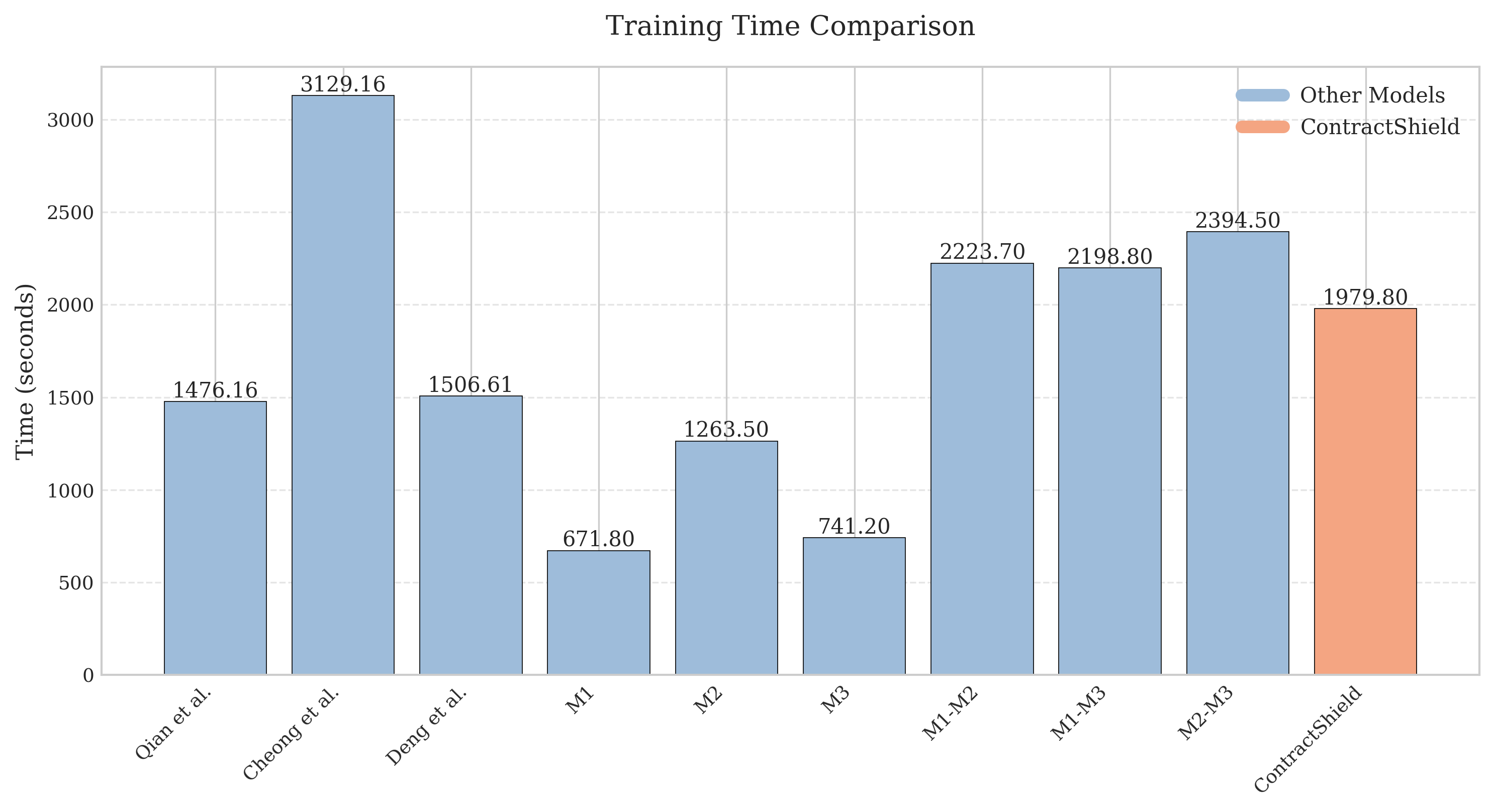}
  \caption{Training time comparison of different methods on the SoliAudit-SmartBugs dataset.}
  \label{fig:training_time}
\end{figure}

\begin{figure}[!ht]
  \centering
  \includegraphics[width=1.0\linewidth]{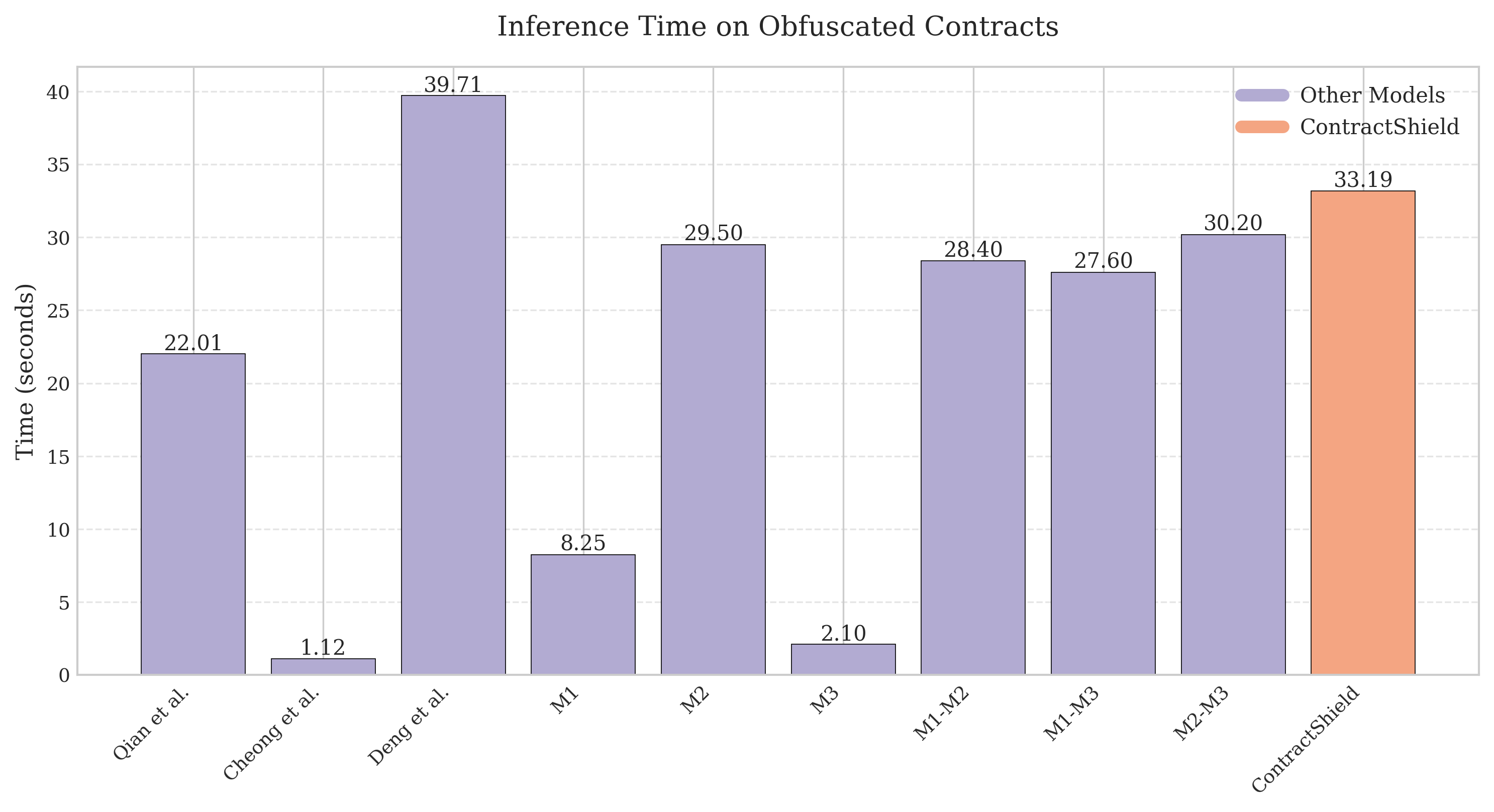}
  \caption{Inference time comparison of different methods on the Test set A.}
  \label{fig:inference_time}
\end{figure}

In this part, we evaluate the training and inference efficiency of our models, including unimodal, bimodal, and multimodal (ContractShield) configurations, alongside the three baseline methods. \textbf{Figure \ref{fig:training_time}} reports training time measured over 18,671 contracts, and all inference times shown in \textbf{Figure \ref{fig:inference_time}} represent the total processing time on Test set A, containing 4,668 contracts, as described in \textbf{Section \ref{sec:datasets}}.


\paragraph{Training efficiency}
\textbf{Figure~\ref{fig:training_time}} shows the training time for all model configurations on the SoliAudit-SmartBugs dataset. Among unimodal models, M3 is the most efficient with a training time of 741.2 seconds, followed by M1 at 671.8 seconds and M2 at 1263.5 seconds, confirming that models trained on a single feature require the least computational resources. As expected, combining two modalities increases training time: M1 combined with M2 requires 2223.7 seconds, M1 with M3 requires 2198.8 seconds, and M2 with M3 requires 2394.5 seconds, with an average of 2272.3 seconds. Despite using hierarchical fusion similar to the bimodal configurations, ContractShield completes training in 1979.8 seconds, approximately 13\% faster than the bimodal average. This indicates that adding a third modality may help stabilize optimization, allowing the fusion hierarchy to converge earlier without increasing computational cost.

Compared with prior works, the differences are more pronounced. Cheong et al.’s GNN-based approach, although unimodal, shows the highest training time at 3129.16 seconds. This is likely due to the computational overhead of message passing across large CFGs, where numerous nodes and edges must be processed. In contrast, Qian et al.’s BLSTM-ATT and Deng et al.’s multimodal model train faster, completing in 1476 and 1506 seconds, respectively. While Deng et al. also incorporate CFG features, their concatenation-based fusion strategy with minimal interaction between modalities keeps computational cost lower than graph-centric approaches like Cheong et al.’s.

Overall, ContractShield achieves a moderate training time compared with other models while providing superior performance, as demonstrated by its robust multi-label detection across three diverse datasets and its resilience to obfuscation techniques.

\paragraph{Inference performance}
\textbf{Figure~\ref{fig:inference_time}} presents the inference time of all model configurations on Test set A. Among unimodal models, M3 is the most efficient with an inference time of 2.10 seconds, followed by M1 at 8.25 seconds and M2 at 29.50 seconds. As expected, combining two modalities increases inference time: M1–M2, M1–M3, and M2–M3 require 28.40, 27.60, and 30.20 seconds, respectively.

When compared to prior works, a notable contrast emerges with training time. Cheong et al.’s GNN-based model, despite its heavy training requirements, achieves the fastest runtime at 1.12 seconds per sample, benefiting from a lightweight evaluation stage. Qian et al.’s BLSTM-ATT takes 22.01 seconds, while Deng et al.’s multimodal model is slower at 39.71 seconds. ContractShield, although taking 33.19 seconds per sample and slightly slower than some other models, offers a reasonable trade-off due to its three-modality design and hierarchical fusion, which enhance predictive accuracy and robustness.
\section{Discussion} \label{sect_Discussion}

\subsection{Discussion on findings}
The experimental findings provide several key insights into the effectiveness and robustness of the proposed ContractShield framework. Overall, the results confirm that hierarchical cross-modal fusion substantially enhances the ability of multimodal models to detect complex and obfuscated vulnerabilities in Ethereum smart contracts.

\textbf{Interpretation of results.} 
The comparative experiments on four datasets (SoliAudit, SmartBugs, CGT Weakness, and DAppScan) demonstrate that ContractShield consistently outperforms both unimodal and existing multimodal baselines. The superior F1-score and Hamming Score indicate that integrating semantic, sequential, and structural modalities through hierarchical attention enables the model to recover complementary cues even when individual modalities are distorted. This robustness stems from the adaptive weighting mechanism, which dynamically balances the contributions of different modalities according to their reliability under obfuscation.

\textbf{Impact of hierarchical fusion.} 
Unlike conventional fusion approaches based solely on concatenation, our design integrates three progressive layers comprising self-attention, cross-modal attention, and adaptive weighting, which collectively capture both intra-modal and inter-modal dependencies. This structure enables the model to retain semantic invariants while alleviating the influence of modality imbalance. The experiments under obfuscation (BiAn and BOSC) further demonstrate that ContractShield sustains strong performance with only minor degradation, confirming the model’s resilience to adversarial transformations that distort structural or syntactic consistency.

\textbf{Practical implications.} 
In practical settings, ContractShield can operate as a core component of automated auditing pipelines or as part of development-time tooling that assists security analysts in detecting complex multifaceted vulnerabilities. Its multimodal architecture supports flexible deployment, whether using source code during development or relying on bytecode and CFGs in closed-source auditing scenarios. Moreover, its comparatively efficient inference time relative to other multimodal systems highlights its potential for semi-real-time vulnerability screening within continuous integration environments.

\textbf{Research significance.} 
The findings highlight the broader implication that cross-modal correlation learning can bridge semantic–structural gaps in code intelligence tasks beyond smart contract analysis. The hierarchical fusion principle could be generalized to other security-critical domains, such as binary vulnerability detection, malware classification, and cross-language code similarity analysis. This underscores ContractShield’s potential as both a methodological and practical contribution to multimodal deep learning for software security.

\subsection{Threats to validity and limitations}

Although the experimental results demonstrate the robustness and superiority of ContractShield, several threats to validity should be recognized.

\textbf{Internal validity.} 
The internal validity of our experiments may be influenced by data quality and preprocessing. Despite extensive cleaning, the benchmark datasets (e.g., SoliAudit, SmartBugs-Wilds, CGT Weakness, and DAppScan) may still contain mislabelled or inconsistent vulnerability annotations that can bias the training process. Moreover, the applied obfuscation tools, BiAn and BOSC, represent controlled transformations that may not encompass the full diversity of adversarial code obfuscation strategies found in the wild. These factors could lead to a slight overestimation of the model’s robustness under real deployment conditions.

\textbf{External validity.} 
Our study primarily focuses on Ethereum smart contracts written in Solidity. Therefore, the results may not generalize to other blockchain ecosystems (e.g., BNB Chain, EOS, or Solana) or to contracts. Additionally, the hierarchical fusion mechanism may introduce complex interdependencies among modalities, making it challenging to generalize. The datasets might not fully reflect the evolving coding styles, compiler updates, and obfuscation patterns in production-level decentralized applications.

\textbf{Construct validity.} 
The evaluation relies on standard metrics such as Hamming Score and F1-score, which may not fully capture the real-world security impact of classification errors. For instance, missing a high-severity vulnerability has greater implications than misclassifying a minor one, yet both contribute equally to the metrics. In addition, the hierarchical fusion mechanism may introduce complex interdependencies among modalities, making it difficult to isolate the contribution of each modality to the final prediction and potentially complicating interpretability.

\textbf{Conclusion validity.} 
The comparison among baseline models might be affected by differences in architectural complexity and computational requirements. While we kept the training conditions consistent across experiments, disparities in model capacity (e.g., parameter size of transformers versus graph neural networks) could influence both accuracy and training efficiency. Therefore, future replication with unified resource constraints and larger-scale cross-validation is encouraged to confirm the observed performance trends.

Overall, these factors represent potential sources of bias and limitations that may influence the interpretation of the reported results. Nevertheless, they do not undermine the general conclusion that hierarchical cross-modal fusion significantly enhances robustness and detection accuracy in multimodal vulnerability analysis.

\section{Conclusion and Future Work} \label{sec:conclusion}

In conclusion, we present ContractShield, a multimodal framework for multi-label vulnerability detection in Ethereum smart contracts that integrates a hierarchical fusion mechanism to combine multiple code representations. The framework systematically integrates three complementary contract representations, including Solidity source code, opcode sequences, and bytecode-derived control flow graphs, to capture both semantic and structural information. At the core of ContractShield is a hierarchical fusion mechanism that combines intra-modal refinement through self-attention, pairwise cross-modal interactions, and adaptive weighting, enabling the model to exploit complementary patterns across modalities. Extensive evaluation on four benchmark datasets, SoliAudit combined with SmartBugs, CGT Weakness, and DAppScan, demonstrates that ContractShield consistently surpasses state-of-the-art unimodal and multimodal models as well as conventional baseline approaches in detecting diverse vulnerability types.

Moreover, our results demonstrate that ContractShield not only maintains robust performance under source-level and bytecode-level obfuscation but also generates more informative and comprehensive feature representations. The hierarchical fusion mechanism allows the model to capture fine-grained intra-modal context, leverage cross-modal interactions, and dynamically adjust attention through task-driven adaptive weighting. This capability reduces the impact of adversarial manipulations aimed at concealing vulnerabilities and enables the model to identify subtle and distributed patterns across multiple code representations. By integrating these complementary insights, ContractShield achieves a clear advantage over simpler concatenation or late fusion strategies and generalizes effectively to diverse datasets and real-world scenarios where smart contracts are deliberately obfuscated or intentionally altered.

For future work, we aim to enhance ContractShield by incorporating additional feature representations, such as intermediate representations (IRs), and integrating abstract syntax trees (ASTs), call graphs (CGs), and data flow graphs (DFGs) into a heterogeneous graph structure. This combined representation will capture more complex relationships, potentially improving both accuracy and robustness, following promising results from related work \cite{nguyen2023mando, luo2024scvhunter}. We also plan to extend ContractShield to detect a broader range of vulnerability types, enhancing its utility across a wider spectrum of smart contract security threats.

\section*{Acknowledgment}

This research is funded by Vietnam National University HoChiMinh City (VNU-HCM) under grant number NCM2025-26-01.

Phan The Duy was funded by the Postdoctoral Scholarship Programme of Vingroup Innovation Foundation (VINIF), VinUniversity, code VINIF.2025.STS.20.



\bibliographystyle{elsarticle-num}
\bibliography{refs}

@article{hu2025enabling,
  title={Enabling Generalized Zero-Shot Vulnerability Classification},
  author={Hu, Jinghao and Guo, Jinsong and Luo, Chen and Hu, Yang and Lanzinger, Matthias and Li, Zhanshan},
  journal={IEEE Transactions on Dependable and Secure Computing},
  year={2025},
  publisher={IEEE}
}

@article{duy2025vulnsense,
  title={Vulnsense: Efficient vulnerability detection in ethereum smart contracts by multimodal learning with graph neural network and language model},
  author={Duy, Phan The and Khoa, Nghi Hoang and Quyen, Nguyen Huu and Trinh, Le Cong and Kien, Vu Trung and Hoang, Trinh Minh and Pham, Van-Hau},
  journal={International Journal of Information Security},
  volume={24},
  number={1},
  pages={48},
  year={2025},
  publisher={Springer}
}

@article{khodadadi2023hymo,
  title={Hymo: Vulnerability detection in smart contracts using a novel multi-modal hybrid model},
  author={Khodadadi, Mohammad and Tahmoresnezhad, Jafar},
  journal={arXiv preprint arXiv:2304.13103},
  year={2023}
}

@inproceedings{bert,
    title = "{BERT}: Pre-training of Deep Bidirectional Transformers for Language Understanding",
    author = "Devlin, Jacob  and
      Chang, Ming-Wei  and
      Lee, Kenton  and
      Toutanova, Kristina",
    booktitle = "Proceedings of the 2019 Conference of the North {A}merican Chapter of the Association for Computational Linguistics: Human Language Technologies, Volume 1 (Long and Short Papers)",
    month = jun,
    year = "2019",
    address = "Minneapolis, Minnesota",
    publisher = "Association for Computational Linguistics",
    doi = "10.18653/v1/N19-1423",
    pages = "4171--4186"
}

@inproceedings{ferreira2020smartbugs,
  title={SmartBugs: A framework to analyze solidity smart contracts},
  author={Ferreira, Jo{\~a}o F and Cruz, Pedro and Durieux, Thomas and Abreu, Rui},
  booktitle={Proceedings of the 35th IEEE/ACM International Conference on Automated Software Engineering},
  pages={1349--1352},
  year={2020}
}

@misc{opcode, url={https://ethereum.org/en/developers/docs/evm/opcodes/}}

@ARTICLE{qian_auto_squen_mol,
  author={Qian, Peng and Liu, Zhenguang and He, Qinming and Zimmermann, Roger and Wang, Xun},
  journal={IEEE Access}, 
  title={Towards Automated Reentrancy Detection for Smart Contracts Based on Sequential Models}, 
  year={2020},
  volume={8},
  number={},
  pages={19685-19695},
  doi={10.1109/ACCESS.2020.2969429}}

@article{li2023smart,
  title={A Smart Contract Vulnerability Detection Method Based on Multimodal Feature Fusion and Deep Learning},
  author={Li, Jinggang and Lu, Gehao and Gao, Yulian and Gao, Feng},
  journal={Mathematics},
  volume={11},
  number={23},
  pages={4823},
  year={2023},
  publisher={MDPI}
}

@article{deng2023smart,
  title={Smart contract vulnerability detection based on deep learning and multimodal decision fusion},
  author={Deng, Weichu and Wei, Huanchun and Huang, Teng and Cao, Cong and Peng, Yun and Hu, Xuan},
  journal={Sensors},
  volume={23},
  number={16},
  pages={7246},
  year={2023},
  publisher={MDPI}
}

@article{zhang2022cbgru,
  title={Cbgru: A detection method of smart contract vulnerability based on a hybrid model},
  author={Zhang, Lejun and Chen, Weijie and Wang, Weizheng and Jin, Zilong and Zhao, Chunhui and Cai, Zhennao and Chen, Huiling},
  journal={Sensors},
  volume={22},
  number={9},
  pages={3577},
  year={2022},
  publisher={MDPI}
}

@inproceedings{vu2023enhancing,
  title={Enhancing Multi-Label Vulnerability Detection of Smart Contract Using Language Model},
  author={Vu, Duong and Nguyen, Tuan and Tong, Van and Souihil, Sami},
  booktitle={2023 5th Conference on Blockchain Research \& Applications for Innovative Networks and Services (BRAINS)},
  pages={1--4},
  year={2023},
  organization={IEEE}
}

@article{tong2024enhancing,
  title={Enhancing BERT-Based Language Model for Multi-label Vulnerability Detection of Smart Contract in Blockchain},
  author={Tong, Van and Dao, Cuong and Tran, Hai-Anh and Tran, Truong X and Souihi, Sami},
  journal={Journal of Network and Systems Management},
  volume={32},
  number={3},
  pages={63},
  year={2024},
  publisher={Springer}
}

@article{yuan2023optimizing,
  title={Optimizing smart contract vulnerability detection via multi-modality code and entropy embedding},
  author={Yuan, Dawei and Wang, Xiaohui and Li, Yao and Zhang, Tao},
  journal={Journal of Systems and Software},
  volume={202},
  pages={111699},
  year={2023},
  publisher={Elsevier}
}

@article{vidal2024vulnerability,
  title={Vulnerability detection techniques for smart contracts: A systematic literature review},
  author={Vidal, Fernando Richter and Ivaki, Naghmeh and Laranjeiro, Nuno},
  journal={Journal of Systems and Software},
  pages={112160},
  year={2024},
  publisher={Elsevier}
}

@article{chu2023survey,
  title={A survey on smart contract vulnerabilities: Data sources, detection and repair},
  author={Chu, Hanting and Zhang, Pengcheng and Dong, Hai and Xiao, Yan and Ji, Shunhui and Li, Wenrui},
  journal={Information and Software Technology},
  volume={159},
  pages={107221},
  year={2023},
  publisher={Elsevier}
}

@inproceedings{kalra2018zeus,
  title={Zeus: analyzing safety of smart contracts.},
  author={Kalra, Sukrit and Goel, Seep and Dhawan, Mohan and Sharma, Subodh},
  booktitle={Ndss},
  pages={1--12},
  year={2018}
}

@inproceedings{luo2024scvhunter,
  title={Scvhunter: Smart contract vulnerability detection based on heterogeneous graph attention network},
  author={Luo, Feng and Luo, Ruijie and Chen, Ting and Qiao, Ao and He, Zheyuan and Song, Shuwei and Jiang, Yu and Li, Sixing},
  booktitle={Proceedings of the IEEE/ACM 46th International Conference on Software Engineering},
  pages={1--13},
  year={2024}
}

@article{zheng2024dappscan,
  title={Dappscan: building large-scale datasets for smart contract weaknesses in dapp projects},
  author={Zheng, Zibin and Su, Jianzhong and Chen, Jiachi and Lo, David and Zhong, Zhijie and Ye, Mingxi},
  journal={IEEE Transactions on Software Engineering},
  year={2024},
  publisher={IEEE}
}

@article{feng2020codebert,
  title={Codebert: A pre-trained model for programming and natural languages},
  author={Feng, Zhangyin and Guo, Daya and Tang, Duyu and Duan, Nan and Feng, Xiaocheng and Gong, Ming and Shou, Linjun and Qin, Bing and Liu, Ting and Jiang, Daxin and others},
  journal={arXiv preprint arXiv:2002.08155},
  year={2020}
}

@inproceedings{di2023consolidation,
  title={Consolidation of ground truth sets for weakness detection in smart contracts},
  author={Di Angelo, Monika and Salzer, Gernot},
  booktitle={International Conference on Financial Cryptography and Data Security},
  pages={439--455},
  year={2023},
  organization={Springer}
}

@inproceedings{jiang2018contractfuzzer,
  title={Contractfuzzer: Fuzzing smart contracts for vulnerability detection},
  author={Jiang, Bo and Liu, Ye and Chan, Wing Kwong},
  booktitle={Proceedings of the 33rd ACM/IEEE international conference on automated software engineering},
  pages={259--269},
  year={2018}
}

@article{beck2024xlstm,
  title={xLSTM: Extended Long Short-Term Memory},
  author={Beck, Maximilian and P{\"o}ppel, Korbinian and Spanring, Markus and Auer, Andreas and Prudnikova, Oleksandra and Kopp, Michael and Klambauer, G{\"u}nter and Brandstetter, Johannes and Hochreiter, Sepp},
  journal={arXiv preprint arXiv:2405.04517},
  year={2024}
}

@inproceedings{contro2021ethersolve,
  title={Ethersolve: Computing an accurate control-flow graph from ethereum bytecode},
  author={Contro, Filippo and Crosara, Marco and Ceccato, Mariano and Dalla Preda, Mila},
  booktitle={2021 IEEE/ACM 29th International Conference on Program Comprehension (ICPC)},
  pages={127--137},
  year={2021},
  organization={IEEE}
}

@inproceedings{cheong2024gnn,
  title={GNN-based Ethereum Smart Contract Multi-Label Vulnerability Detection},
  author={Cheong, Yoo-Young and Shin, Jihwan and Kim, Taekyung and Ahn, Jinhyun and Im, Dong-Hyuk and others},
  booktitle={2024 International Conference on Information Networking (ICOIN)},
  pages={57--61},
  year={2024},
  organization={IEEE}
}

@inproceedings{nguyen2023mando,
  title={Mando-hgt: Heterogeneous graph transformers for smart contract vulnerability detection},
  author={Nguyen, Hoang H and Nguyen, Nhat-Minh and Xie, Chunyao and Ahmadi, Zahra and Kudendo, Daniel and Doan, Thanh-Nam and Jiang, Lingxiao},
  booktitle={2023 IEEE/ACM 20th International Conference on Mining Software Repositories (MSR)},
  pages={334--346},
  year={2023},
  organization={IEEE}
}

@article{grech2018madmax,
  title={Madmax: Surviving out-of-gas conditions in ethereum smart contracts},
  author={Grech, Neville and Kong, Michael and Jurisevic, Anton and Brent, Lexi and Scholz, Bernhard and Smaragdakis, Yannis},
  journal={Proceedings of the ACM on Programming Languages},
  volume={2},
  number={OOPSLA},
  pages={1--27},
  year={2018},
  publisher={ACM New York, NY, USA}
}

@misc{gemini_dao_hack,
  author       = {{Gemini Cryptopedia}},
  title        = {The DAO: What Was the DAO Hack?},
  url          = {https://www.gemini.com/cryptopedia/the-dao-hack-makerdao},
  note         = {Accessed: April 12, 2025}
}

@misc{wanjiku2023dforce,
  author       = {Samuel Mbaki Wanjiku},
  title        = {dForce confirms the return of exploited \$3.65M to their vaults},
  year         = {2023},
  url          = {https://crypto.news/dforce-confirms-the-return-of-exploited-3-65m-to-their-vaults/},
  note         = {Accessed: April 12, 2025}
}

@article{zhang2023bian,
  title={Bian: Smart contract source code obfuscation},
  author={Zhang, Pengcheng and Yu, Qifan and Xiao, Yan and Dong, Hai and Luo, Xiapu and Wang, Xiao and Zhang, Meng},
  journal={IEEE Transactions on Software Engineering},
  volume={49},
  number={9},
  pages={4456--4476},
  year={2023},
  publisher={IEEE}
}

@inproceedings{zhang2020source,
  title={Source code obfuscation for smart contracts},
  author={Zhang, Meng and Zhang, Pengcheng and Luo, Xiapu and Xiao, Feng},
  booktitle={2020 27th Asia-Pacific Software Engineering Conference (APSEC)},
  pages={513--514},
  year={2020},
  organization={IEEE}
}

@inproceedings{upadhya2024quadracode,
  title={QuadraCode AI: Smart Contract Vulnerability Detection with Multimodal Representation},
  author={Upadhya, Jiblal and Upadhyay, Kritagya and Sainju, Arpan and Poudel, Samir and Hasan, Md Nahid and Poudel, Khem and Ranganathan, Jaishree},
  booktitle={2024 33rd International Conference on Computer Communications and Networks (ICCCN)},
  pages={1--9},
  year={2024},
  organization={IEEE}
}

@article{wu2024comprehensive,
  title={A comprehensive survey of smart contract security: State of the art and research directions},
  author={Wu, Guangfu and Wang, HaiPing and Lai, Xin and Wang, Mengmeng and He, Daojing and Chan, Sammy},
  journal={Journal of Network and Computer Applications},
  pages={103882},
  year={2024},
  publisher={Elsevier}
}

@article{jie2023novel,
  title={A novel extended multimodal AI framework towards vulnerability detection in smart contracts},
  author={Jie, Wanqing and Chen, Qi and Wang, Jiaqi and Koe, Arthur Sandor Voundi and Li, Jin and Huang, Pengfei and Wu, Yaqi and Wang, Yin},
  journal={Information Sciences},
  volume={636},
  pages={118907},
  year={2023},
  publisher={Elsevier}
}

@article{lian2024universal,
  title={A Universal and Efficient Multi-modal Smart Contract Vulnerability Detection Framework for Big Data},
  author={Lian, Wenjuan and Bao, Zikang and Zhang, Xinze and Jia, Bin and Zhang, Yang},
  journal={IEEE Transactions on Big Data},
  year={2024},
  publisher={IEEE}
}

@inproceedings{yu2022bytecode,
  title={Bytecode obfuscation for smart contracts},
  author={Yu, Qifan and Zhang, Pengcheng and Dong, Hai and Xiao, Yan and Ji, Shunhui},
  booktitle={2022 29th Asia-Pacific Software Engineering Conference (APSEC)},
  pages={566--567},
  year={2022},
  organization={IEEE}
}

@inproceedings{liao2019soliaudit,
  title={Soliaudit: Smart contract vulnerability assessment based on machine learning and fuzz testing},
  author={Liao, Jian-Wei and Tsai, Tsung-Ta and He, Chia-Kang and Tien, Chin-Wei},
  booktitle={2019 Sixth International Conference on Internet of Things: Systems, Management and Security (IOTSMS)},
  @pages={458--465},
  year={2019},
  organization={IEEE}
}

@article{li2025interaction,
  title={Interaction-aware vulnerability detection in smart contract bytecodes},
  author={Li, Wenkai and Li, Xiaoqi and Mao, Yingjie and Zhang, Yuqing},
  journal={IEEE Transactions on Dependable and Secure Computing},
  year={2025},
  publisher={IEEE}
}

@article{yang2026byteeye,
  title={ByteEye: A smart contract vulnerability detection framework at bytecode level with graph neural networks},
  author={Yang, Jinni and Liu, Shuang and Dai, Surong and Fang, Yaozheng and Xie, Kunpeng and Lu, Ye},
  journal={Automated Software Engineering},
  volume={33},
  number={1},
  pages={1--38},
  year={2026},
  publisher={Springer}
}

@article{crisostomo2025machine,
  title={Machine learning methods for detecting smart contracts vulnerabilities within Ethereum blockchain- A review},
  author={Crisostomo, Joao and Bacao, Fernando and Lobo, Victor},
  journal={Expert Systems with Applications},
  pages={126353},
  year={2025},
  publisher={Elsevier}
}

@article{bresil2025deep,
  title={Deep Learning-based Vulnerability Detection Solutions in Smart Contracts: A Comparative and Meta-Analysis of Existing Approaches},
  author={Bresil, Michael and Prasad, Pwc and Sayeed, Md Shohel and Bukar, Umar Ali},
  journal={IEEE Access},
  year={2025},
  publisher={IEEE}
}

@article{wei2025advanced,
  title={Advanced smart contract vulnerability detection via llm-powered multi-agent systems},
  author={Wei, Zhiyuan and Sun, Jing and Sun, Yuqiang and Liu, Ye and Wu, Daoyuan and Zhang, Zijian and Zhang, Xianhao and Li, Meng and Liu, Yang and Li, Chunmiao and others},
  journal={IEEE Transactions on Software Engineering},
  year={2025},
  publisher={IEEE}
}

@article{balci2025examining,
  title={Examining the effectiveness of transformer-based smart contract vulnerability scan},
  author={Balci, Emre and Aydede, Timucin and Yilmaz, Gorkem and Soyak, Ece Gelal},
  journal={Journal of Systems and Software},
  pages={112593},
  year={2025},
  publisher={Elsevier}
}

@article{ding2025smartguard,
  title={SmartGuard: An LLM-enhanced framework for smart contract vulnerability detection},
  author={Ding, Hao and Liu, Yizhou and Piao, Xuefeng and Song, Huihui and Ji, Zhenzhou},
  journal={Expert Systems with Applications},
  volume={269},
  pages={126479},
  year={2025},
  publisher={Elsevier}
}

@inproceedings{upadhya2024vulnfusion,
  title={VulnFusion: Exploiting multimodal representations for advanced smart contract vulnerability detection},
  author={Upadhya, Jiblal and Sainju, Arpan and Upadhyay, Kritagya and Poudel, Samir and Hasan, Md Nahid and Poudel, Khem and Ranganathan, Jaishree},
  booktitle={2024 6th International Conference on Blockchain Computing and Applications (BCCA)},
  pages={505--515},
  year={2024},
  organization={IEEE}
}

@inproceedings{sahu2021adaptive,
  title={Adaptive fusion techniques for multimodal data},
  author={Sahu, Gaurav and Vechtomova, Olga},
  booktitle={Proceedings of the 16th conference of the European chapter of the Association for Computational Linguistics: Main Volume},
  pages={3156--3166},
  year={2021}
}

@article{cai2024fine,
  title={Fine-grained smart contract vulnerability detection by heterogeneous code feature learning and automated dataset construction},
  author={Cai, Jie and Li, Bin and Zhang, Tao and Zhang, Jiale and Sun, Xiaobing},
  journal={Journal of Systems and Software},
  volume={209},
  pages={111919},
  year={2024},
  publisher={Elsevier}
}

@article{wang2025review,
  title={A Review of Learning-based Smart Contract Vulnerability Detection: A Perspective on Code Representation},
  author={Wang, Ben and Tong, Yanxiang and Ji, Shunhui and Dong, Hai and Luo, Xiapu and Zhang, Pengcheng},
  journal={ACM Transactions on Software Engineering and Methodology},
  year={2025},
  publisher={ACM New York, NY}
}

@article{HAN2025130619,
    title = {MKDD-Vul: A Lightweight Multi-modal Knowledge Distillation Framework for Detecting Vulnerabilities in Smart Contracts},
    journal = {Expert Systems with Applications},
    pages = {130619},
    year = {2025},
    issn = {0957-4174},
    doi = {https://doi.org/10.1016/j.eswa.2025.130619},
    url = {https://www.sciencedirect.com/science/article/pii/S0957417425042344},
    author = {Daojun Han and Pan Qi and Juntao Zhang and Ziliang Guo and Linkun Fan},
}

@article{wang2025tmf,
  title={TMF-Net: Multimodal smart contract vulnerability detection based on multiscale transformer fusion},
  author={Wang, Tengfei and Zhao, Xiangfu and Zhang, Jiarui},
  journal={Information Fusion},
  volume={122},
  pages={103189},
  year={2025},
  publisher={Elsevier}
}





\end{document}